\documentclass[pre,11pt, onecolumn,amssymb,amsmath,showpacs]{revtex4-1}

\usepackage[margin=2cm]{geometry}
\usepackage{graphicx}
\usepackage{epsfig}
\usepackage{latexsym}
\usepackage{bm}
\usepackage{ulem}
\usepackage{physics}

\usepackage{amsmath}
\usepackage{amsfonts}
\usepackage{amssymb}
\usepackage{float}

\usepackage{color}
\definecolor{dgreen}{rgb}{0,0.7,0}

\begin{document}

\title{Gap statistics of two interacting run and tumble particles in one dimension}
\author{Arghya Das, Abhishek Dhar and Anupam Kundu} 
\affiliation{International Centre for Theoretical Sciences, Tata Institute of Fundamental Research, Bengaluru 560089, India}


\begin{abstract}
\noindent
We study the dynamics of the separation (gap) between a pair of interacting run and tumble particles (RTPs) moving in one dimension in the presence of additional thermal noise. On a ring geometry the distribution of the gap approaches a steady state.  We analytically compute this distribution and find that this is exponentially localised in space, in contrast to the `jammed' configuration, seen earlier in the absence of thermal noise. We also study the relaxation which is an exponential, characterised by a time scale $\tau_r$. We observe that this time scale undergoes a crossover from a size independent value to a size dependent form with increasing size $l$ of the ring. We study the full eigenvalue spectrum of the evolution operator $\mathcal{L}$ and find that the spectrum can be classified into four sectors depending on the symmetries of $\mathcal{L}$. For large $l$, we find explicit expressions for the low lying eigenvalues in each of the sectors.
On infinite line the separation does not reach a steady state. In the long times we find that the particles behave as interacting Brownian particles, except for the presence of a peak in the distribution at small separation which is a remnant of activity.
\end{abstract}

\maketitle

\section{Introduction}
\noindent
Self-propelled particles (SPPs), also known as `active' particles, have gained widespread interest owing to their applicability in vast range of nonequilibrium phenomena \cite{Ramaswamy2017,Galajda,Schwarz,Popkin,Vicsek,Nedelec,Prost,Palacci}. SPPs violate microscopic reversibility at the level of individual constituents \cite{Battle,Burkholder} and exhibit remarkable collective behaviour including flocking \cite{Popkin,Vicsek,Biplab}, segregation of active and passive particles \cite{Stenhammar,Pritha}, or of active particles with opposite chirality from a mixture \cite{Volpe}, non-Boltzmann distribution in the steady state \cite{Bechinger, Takatori, Szamel, Walsh, Dauchot}, Casimir effect \cite{Ni,Ray}, `active' pressure \cite{Solon-pressure,Duzgun,Fily}, motility induced phase separation and clustering \cite{Redner, Cates, Theers}, emergent ordering and pattern formation \cite{Rakesh-transition, Rakesh-dynamics, Schaller, Edwards}, giant number fluctuation \cite{Toner-Tu,Menon,Fily-Marchetti}, etc.
In literature several active particle models have been devised, the simplest and arguably most widely studied variants being the active Brownian particles (ABPs) and run and tumble particles (RTPs). On the microscopic level it is known that even a single active particle can demonstrate remarkable nontrivial behaviour, many of which have been analytically characterised recently \cite{Pototsky, UB, one-RT-particle, Kanaya-ABP, confined-RTP}. Theoretical approaches in lines of (mostly phenomenological) hydrodynamics have been proposed for interacting active systems to quantify behaviour like macroscopic phase separation and ordering \cite{Ramaswamy, Toner, Bolley, Bertin-nematic, Solon, Puglisi, Kroy, Tailleur}. However, a systematic and comprehensive \textit{microscopic} understanding of these properties has not yet been achieved.

In a recent work, Slowman {\it et.~al.} \cite{Slowman_prl2016} considered a pair of RTPs moving on a periodic (circular) lattice and interacting via nearest neighbour exclusion. Their motivation was to understand how jamming, clustering and effective attractive interaction emerge from a microscopic point of view.  They show that this simple dynamics leads to a steady state with a `bound configuration' characterised by an exponentially decaying component and a `jammed' part where the particles are stuck at neighbouring sites with finite probability.
They suggested that this `jammaing' feature can be understood to arise from an {\it effective attraction} between a pair of {\it passive particles} which serves as a possible explanation of self assembly in the many particle case.

However, in their study they do not consider the effect of thermal noise which is ubiquitous in real systems, such as the interior of cell, bacterial colony etc. Moreover,  recent studies suggest that introduction of finite diffusion in the dynamics is necessary for macroscopic phase separation in SPPs \cite{Levis}. Hence in order to have a proper microscopic description of more realistic SPPs one needs to consider thermal diffusion into account in addition to the usual persistent dynamics. 

Furthermore, most of the examples of real active particles such as bacteria, Janus particles etc. move on continuous space. Hence it is also important to consider model systems on continuous space. If the model considered in \cite{Slowman_prl2016} is studied on a continuous ring instead of a circular lattice then one finds that in the steady state the particles are either stuck together (characterised by a Dirac delta) or moving independently (characterised by a uniform distribution of gaps between the two particles). This implies that there is no nontrivial length scale in the `bound' configurations. 
In \cite{Slowman_tumble-duration} it has been shown that if the particles take a finite duration to tumble then the `bound' component (in addition to a `jammed' component) in the steady state posses a finite length scale. 
Here we show that addition of thermal noise (\emph{i.e.} diffusion) 
also generates `bound' states that decay exponentially over a finite length scale.

In this paper, we studied the dynamics of two interacting RTPs in presence of external thermal noise in one dimension both on ring geometry and on infinite line. We denote the positions of these two point particles at time $t$ by $x_1(t),~x_2(t)$, respectively. Each particle moves with speed $v$ in either positive or negative direction denoted by `spin' variables $\sigma_{1},~\sigma_{2}$ that can take values $\pm 1$. These spins change sign independently with rate $\omega$ and thus are described by  telegraphic processes denoted by $\sigma_1(t)$ and $\sigma_2(t)$. In addition, the particles interact by hard core repulsion. The equation of motion of the two particles are given by
\begin{align}
\begin{split}
\frac{d x_1}{dt}=v \sigma_1(t)+\eta_1(t) +F_c~, \\
\frac{d x_2}{dt}=v \sigma_2(t)+\eta_2(t) -F_c~, 
\end{split}
\label{eom_12}
\end{align}
where  $\eta_1(t),~\eta_2(t)$  are two independent Gaussian white noises both with zero mean and variance $2D$. The repulsive contact force term $F_c$  prevents the particles from crossing each other and is described later as boundary conditions for the distribution of the separation.
The  dynamics of individual particles thus has four parts: uniform motion in a given direction with speed $v$ (activity), random flipping of the direction at rate $\omega$ (tumble), the usual Brownian fluctuations with diffusion constant $D$ and hardcore repulsion. We note that this model of two  RTPs has recently been studied in \cite{Doussal19} where the authors looked at the first passage problem for the separation between the particles on the infinite line and obtained several analytic results. This paper did not include thermal noise. Moreover, as we will see, the first passage problem requires solving the master equations with absorbing boundary conditions which are different from the ones that we discuss. 

\begin{figure}[h]
\begin{center}
\includegraphics[width=12cm,angle=0.0]{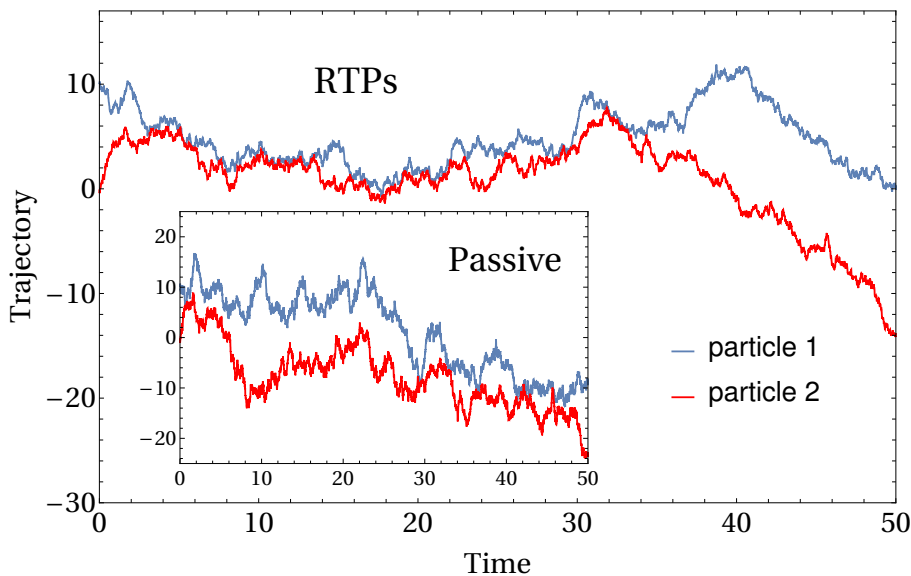}
\end{center}
\caption{Typical trajectories of two RTPs in one dimension with exclusion in presence of additional external thermal noise.  When they come close, they tend to stay together for longer duration compared to two passive Brownian particles (shown in inset) having equivalent diffusion coefficients (discussed later in the text) with exclusion interaction. }
\label{trajectories}
\end{figure}

Typical trajectories of the two particles are shown in Fig.~(\ref{trajectories}) in comparison with the trajectories [Inset of Fig.~(\ref{trajectories})] of two passive Brownian particles. We observe that when the two RTPs come close they tend to stay close for a long duration compared to passive particles which is a manifestation of the presence of activity. This suggests the activity present in the dynamics generates an effective attractive pairwise interaction which can be understood from the dynamics and statistical properties of the separation or gap between the two particles.

In this paper we are mainly interested in the statistics of the gap (or separation) between the two particles which is defined as follows. On the infinite line we assume $x_2(0) < x_1(0)$ and define the gap of the  particles as $x(t)=x_1(t) - x_2(t)$.  On the other hand when the particles move on a ring of length $l$, we  measure the gap $x(t)$ at time $t$ in the clockwise direction from the position $x_2(t)$ of the second particle to the position $x_1(t)$ of the first particle. Thus by definition, in both cases $x(t) \ge 0$.
In contrast to the study of Slowman {\it et.~al.} \cite{Slowman_prl2016}, we observe that in our model defined on continuous space, the presence of thermal diffusion has a dramatic effect on the steady state and relaxation properties of the gap variable $x(t)$. Below we provide a brief summary of our main results along with the organisation of the paper.

In Sec.~\ref{ME-GF} we start by writing the master equations for the probability distribution of the gap (separation) $x$ between the two particles with boundary conditions on the ring geometry as well as on the infinite line. We provide a general formulation for solving these equations in these domains. The detailed discussion  for the ring geometry is presented in Sec.~\ref{1D-continuum-ring}.   In the ring geometry, we find that in the large time limit, the probability distribution of the gap approaches a steady state  that can be expressed as a sum of two exponentially decaying functions with two different length scales representing two types of `bound states'. This result is in contrast with the situation when the two particles are moving on a periodic lattice without diffusion \cite{Slowman_prl2016} where the steady state is described by a single exponentially decaying bound component along with a `jammed' state. This is followed by a discussion on the relaxation to the steady state in Sec.~\ref{continuum-ring-relaxation} where we find that the relaxation rate, given by the second largest eigenvalue $\lambda_1(<0)$ of the evolution operator $\mathcal{L}$ in the master equation, undergoes a `transition' upon increasing the size $l$ of the ring.
Interesting crossover in the relaxation behaviour upon tuning the activity parameters (like the tumble rate), as a function of system size or trap strength, have been reported earlier for both single particle case \cite{confined-RTP, Evans_pre-RT-spectrum}, and for interacting RTPs on a ring \cite{Evans_pre-RT-spectrum}.
Here we show that for given $\omega$ and $D$ there exists a length $l_c\sim \sqrt{D/\omega}$  such that for $l\leq l_c$, the relaxation rate is independent of $l$ and is given by $\lambda_1 = -2\omega$. On the other hand for $l>l_c$, we find that $\lambda_1$ starts decreasing with $l$ and for large $l~(\gg l_c)$ it is approximately given by $\lambda_1 \simeq - 2 \pi^2 D_e/l^2$ where $D_e=D+v^2/(2\omega)$.  Equivalently this transition can also be observed by tuning $\omega$ keeping $l, D$ fixed. In Sec.~\ref{spectrum} we study the full eigenvalue spectrum of the evolution operator $\mathcal{L}$ in the ring geometry. Here we follow the ideas developed in \cite{Evans_pre-RT-spectrum} and 
 show that in the present case the master equations and the associated boundary conditions posses two symmetries, involving spin flips and space reflections, described by operators $\mathcal{O}_s$ and $\mathcal{O}_x$ which commute with $\mathcal{L}$. As a result, diagonalising $\mathcal{L},~\mathcal{O}_s$ and $\mathcal{O}_x$ simultaneously, one can represent the eigenstates by the eigenvalues $\lambda_n,~o_s$ and $o_x$ of these three operators, respectively. Depending on the allowed values $\pm1$ for both  $o_{s}$ and $o_x$\,, we find that it is possible to divide the whole spectrum into four symmetry sectors --- even-even ($o_s=1,~o_x=1$), even-odd ($o_s=1,~o_x=-1$), odd-even ($o_s=-1,~o_x=1$) and odd-odd ($o_s=-1,~o_x=-1$). We show that  all the eigenvalues $\lambda_n$ in the odd-even and odd-odd sectors can be obtained exactly and are given by 
\begin{align}
\lambda_{n}&=-2\omega-\frac{2D\pi^2 (2n)^2}{l^2}~\text{for}~n=0,1,2,~~(\text{odd-even}), \nonumber \\
\lambda_{n}&=-2\omega-\frac{2D\pi^2 (2n+1)^2}{l^2}~\text{for}~n=0,1,2,~ ~(\text{odd-odd}). \nonumber
\end{align} 
The eigenvalues in the other two sectors can be calculated numerically from an explicit equation involving determinant of $3 \times 3$ matrices, with analytic expressions obtainable in the large $l$ limit. Finally, in Sec.~\ref{infinite_line} we study the gap between two particles on the infinite line where, as expected, we find that the system does not reach a steady state in the large time limit. Instead the gap distribution exhibit a scaling behavior similar to the passive case with an effective diffusion constant $D_e$, and some remnants of activity showing up as a peak in the distribution at small separation. In Sec.~\ref{summary} we conclude our paper with a discussion on possible further extensions of our work.

\section{master equations and the general formulation}
\label{ME-GF}
\noindent
From the equations of motion in \eqref{eom_12}, it is easy to observe that the centre of mass, $X(t)=(x_1(t)+x_2(t))/2$, undergoes a `free' RTP motion with an effective telegraphic noise $\tilde{\Sigma}=(\sigma_1+\sigma_2)/2$ and an effective Brownian noise $\tilde{\eta}=(\eta_1+\eta_2)/2$. On the other hand, the gap coordinate,  $x(t)=x_1(t)-x_2(t)$, carries the effect of the interaction. Since we are interested in the dynamics of the gap, we focus on obtaining the master equation for $\mathcal{P}_{\sigma_2\sigma_1}(x,t)$ denoting the probability density of finding the two particles at separation $x$ at time $t$ with spin values $\sigma_1$ and $\sigma_2$ of the two particles. Since each spin can take two values $\pm 1$, there are four components of the probability distribution. The master equation describing their evolution (valid for $x\ge0$) is given by
\begin{align}
\begin{split}
\dot{\mathcal{P}}_{++}&=2D\frac{\partial^2{\mathcal{P}}_{++}}{\partial x^2}+\omega[\mathcal{P}_{+-}+\mathcal{P}_{-+}-2\mathcal{P}_{++}], \\
\dot{\mathcal{P}}_{+-}&=2D\frac{\partial^2{\mathcal{P}}_{+-}}{\partial x^2}+2v\frac{\partial{\mathcal{P}}_{+-}}{\partial x}\\
&~~~ +\omega[{\mathcal{P}}_{++}+{\mathcal{P}}_{--}-2{\mathcal{P}}_{+-}],\\
\dot{\mathcal{P}}_{-+}&=2D\frac{\partial^2{\mathcal{P}}_{-+}}{\partial x^2}-2v\frac{\partial{\mathcal{P}}_{-+}}{\partial x}\\
&~~~~ +\omega[{\mathcal{P}}_{++}+{\mathcal{P}}_{--}-2{\mathcal{P}}_{-+}],\\
\dot{\mathcal{P}}_{--}&=2D\frac{\partial^2{\mathcal{P}}_{--}}{\partial x^2}+\omega[{\mathcal{P}}_{+-}+{\mathcal{P}}_{-+}-2{\mathcal{P}}_{--}].
\end{split}
\label{ME-1d-continuum}
\end{align}
The second order derivative terms on the right hand side are coming from the thermal noises, whereas the first order derivative terms appear due to the drift with speed $v$. The terms proportional to $\omega$ appear due to the changing of sign (tumbling) of the telegraphic noises $\sigma_1(t)$ and $\sigma_2(t)$. 
Note that the marginal gap distribution is given by the sum
\[\mathcal{P}(x,t) = \mathcal{P}_{++}(x,t)+\mathcal{P}_{+-}(x,t)+\mathcal{P}_{-+}(x,t)+\mathcal{P}_{--}(x,t).\]
These equations need to be solved with the appropriate boundary conditions which we now specify.  
On the \textit{ring geometry}, the two sets of boundary conditions at $x=0$ and $x=l$ are obtained by setting the individual probability  currents going out through these boundaries to zero because of hardcore repulsion. Thus, at $x=0$ we get  
 \begin{align}
& \frac{\partial {\mathcal{P}}_{++}(x,t)}{\partial x}\Big{|}_{x=0}=0, \label{BC1-inf-line}\\
D & \frac{\partial {\mathcal{P}}_{+-}(x,t)}{\partial x}\Big{|}_{x=0}+v{\mathcal{P}}_{+-}(0,t)=0,\\
D & \frac{\partial {\mathcal{P}}_{-+}(x,t)}{\partial x}\Big{|}_{x=0}-v{\mathcal{P}}_{-+}(0,t)=0,\\
& \frac{\partial {\mathcal{P}}_{--}(x,t)}{\partial x}\Big{|}_{x=0}=0, \label{BC4-inf-line}
\end{align}
while at $x=l$ we get
\begin{align}
& \frac{\partial {\mathcal{P}}_{++}(x,t)}{\partial x}\Big{|}_{x=l}=0, \label{BC1-ring}\\
D & \frac{\partial {\mathcal{P}}_{+-}(x,t)}{\partial x}\Big{|}_{x=l}+v{\mathcal{P}}_{+-}(l,t)=0,\\
D& \frac{\partial {\mathcal{P}}_{-+}(x,t)}{\partial x}\Big{|}_{x=l}-v{\mathcal{P}}_{-+}(l,t)=0,\\
& \frac{\partial {\mathcal{P}}_{--}(x,t)}{\partial x}\Big{|}_{x=l}=0. \label{BC4-ring}
\end{align}
On the \textit{infinite line}, the boundary conditions at $x=0$ remain the same as given in Eqs.~\eqref{BC1-inf-line}-\eqref{BC4-inf-line} and the boundary condition at $x \to \infty$ are obviously
\begin{equation}
\mathcal{P}_{\sigma_2\sigma_1}(x\rightarrow\infty,t)\rightarrow 0,~~\text{for}~\sigma_{1,2}=\pm1. \label{BC-infinity}
\end{equation}
\noindent
Note that  the first passage problem for the gap, studied in \cite{Doussal19}, leads to the same master equation (without the diffusion terms) but with  different boundary conditions. 

Defining the column vector $|\mathcal{P}(x,t)\rangle \equiv (\mathcal{P}_{++},\mathcal{P}_{+-},\mathcal{P}_{-+},\mathcal{P}_{--})^T$, we note that Eqns.~\eqref{ME-1d-continuum}-\eqref{BC4-ring} can be written in the form $\partial_t  |\mathcal{P}(x,t)\rangle = \mathcal{L} |\mathcal{P}(x,t)\rangle$ where $\mathcal{L}$ is a linear operator. The general time dependent solution can then be written as 
\begin{eqnarray}
|\mathcal{P}(x,t)\rangle = \sum_{n=0}^\infty c_{n} e^{\lambda_n t} |\phi_n(x)\rangle~,\label{transient-distribution}
\end{eqnarray}
where $\lambda_n$ is the $n^{\text{th}}$ eigenvalue (assuming the spectrum to be discrete and ordered \emph{i.e.} $\text{Re}[\lambda_n] \ge \text{Re}[\lambda_{n+1}],~\forall n$) and $|\phi_n \rangle$ the corresponding right eigenfunctions of the operator $\mathcal{L}$, i.e, $\mathcal{L} |\phi_n(x)\rangle = \lambda_n |\phi_n(x)\rangle$. More explicitly,
\begin{eqnarray}
&& 2D\frac{d^2 \phi_n^{++}}{d x^2}+\omega[\phi_n^{+-}+\phi_n^{-+}-2\phi_n^{++}] \label{1d-continuum_1}=\lambda_n \phi^{++}_n, \nonumber\\
&& 2D\frac{d^2 \phi_n^{+-}}{d x^2}+2v\frac{d \phi_n^{+-}}{d x}+\omega[\phi_n^{++}+\phi_n^{--}-2\phi_n^{+-}]=\lambda_n \phi^{+-}_n, \nonumber\\
&& 2D\frac{d^2\phi_n^{-+}}{d x^2}-2v\frac{d \phi_n^{-+}}{d x}+\omega[\phi_n^{++}+\phi_n^{--}-2\phi_n^{-+}]=\lambda_n \phi^{-+}_n,\nonumber\\
&& 2D\frac{d^2\phi_n^{--}}{d x^2}+\omega[\phi_n^{+-}+\phi_n^{-+}-2\phi_n^{--}]=\lambda_n \phi^{--}_n.\nonumber
\end{eqnarray}
We try solutions of the form $|\phi_n(x)\rangle=e^{k_n x} (A_n^{++},A_n^{+-},A_n^{-+},A_n^{--})^T$, where~$A_{n}^{\sigma,\sigma'}$ are constants. When this form is inserted in the above coupled differential equations for $|\phi_n(x)\rangle$, we get the eigenvalue equation, $M(k_n)|A_{n}\rangle = \lambda_n |A_{n}\rangle$, with
\[
 M(k)=(2Dk^2-2\omega){\bf I}+
 \begin{bmatrix}
    0 & \omega & \omega & 0 \\
    \omega & 2vk & 0 & \omega \\
    \omega & 0 & -2vk & \omega\\
    0 & \omega & \omega & 0
  \end{bmatrix}. \label{Markov-matrix}
\]
Setting ${\rm Det}[M-\lambda {\bf I}]=0$ for nonzero solutions for $|A_{\lambda}\rangle$, we get the following  set of relations between $k$'s and $\lambda$,
\begin{eqnarray}
 k^2 =
\left\{
\begin{array}{ll}
&k_a^2=\frac{\omega + \frac{\lambda}{2}}{D} ~~{\rm (doubly~degenerate)}\cr
&k_b^2=\frac{2\omega D + v^2}{2D^2}+\frac{\lambda}{2D}+\frac{\sqrt{(2\omega D+v^2)^2+2\lambda D v^2}}{2D^2}\cr
&k_c^2=\frac{2\omega D + v^2}{2D^2}+\frac{\lambda}{2D}-\frac{\sqrt{(2\omega D+v^2)^2+2\lambda D v^2}}{2D^2}
\end{array}\label{spatial-modes}
\right .~~
\end{eqnarray}
Thus, for a given $\lambda_n$, there are eight allowed values of $k$,
\begin{align}
\begin{split}
&k^{(1)}_n=k_a(\lambda_n),~k^{(2)}_n=-k_a(\lambda_n), \\
&k^{(3)}_n=k_a(\lambda_n),~k^{(4)}_n=-k_a(\lambda_n), \\
&k^{(5)}_n=k_b(\lambda_n),~k^{(6)}_n=-k_b(\lambda_n), \\
&k^{(7)}_n=k_c(\lambda_n),~k^{(8)}_n=-k_c(\lambda_n),
\end{split}
\label{ks}
\end{align}
and the corresponding eigenvectors are
\begin{align}
\begin{split}
|A_n^{(i)}\rangle =   \begin{bmatrix}
   1 \\
   0 \\
   0 \\
   -1
  \end{bmatrix},~{\rm for}~i=1~\text{and}~2~,~
|A_n^{(i)}\rangle = 
  \begin{bmatrix}
   1 \\
   a_2(k^{(i)}_n) \\
   a_3(k^{(i)}_n)\\
   1
  \end{bmatrix} {\rm for}~i=3,\ldots 8,  
  \end{split}
  \label{evectors}
\end{align}
where $a_2(k) = \frac{\omega}{\lambda/2-Dk^2-vk+\omega},~\mbox{and}~ a_3(k) = \frac{\omega}{\lambda/2-D{k}^2+vk+\omega}$. It is worth noting that, $a_2(-k)=a_3(k)$ for all $i=1,2,..,8$. This happens because of the presence of specific symmetries of the problem related to to inversion in both position and spin space (see sec.~\ref{spectrum}).
The eigenvalues $\lambda_n$ are yet undetermined and these will be determined once we impose boundary conditions. To find these, we write our required eigenvector $|\phi_n(x)\rangle$ in the general form 
\begin{align} \label{eigenvec}
|\phi_n(x) \rangle = \sum_{i=1}^8 \alpha_n^{(i)} e^{k_n^{(i)} x} |A_n^{(i)} \rangle,
\end{align}
where the eight constants $\alpha_n^{(i)}$ and the unknown eigenvalues $\lambda_n$ will be determined by the eight boundary conditions. The full time dependent solution is then obtained by evaluating $c_n$ from the initial condition.

We first study the  steady state distribution and then the relaxation properties  on the  ring geometry. The results on infinite line are presented in the subsequent section.

\section{Gap distribution on a ring}
\label{1D-continuum-ring}

\subsection{Steady state}\label{continuum-ring-SS}
\begin{figure}[h]
\begin{center}
\includegraphics[width=12.0cm,angle=0.0]{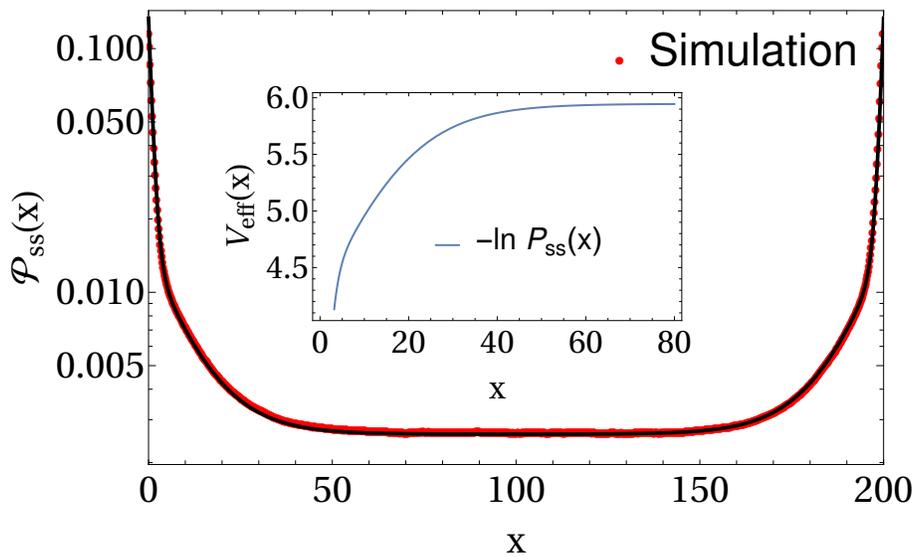}
\end{center}
\caption{Steady state gap distribution, $\mathcal{P}_{ss}(x)$ on continuum ring with parameters : $l=200,~D=1.0,~v=1.0,\omega=0.01$. The effective potential is shown in the inset.}\label{steady-state-fig}
\end{figure}

\noindent
The steady state is given by the eigenfunction corresponding to the largest eigenvalue $\lambda_0=0$. Inserting this value in Eq.~\eqref{spatial-modes} one gets the eight $k$ values using which in Eq.~\eqref{eigenvec} one obtains the steady state $|\mathcal{P}(x)\rangle = |\phi_0(x)\rangle$. The  $\alpha_0^{(i)}$s for $i=1,2,...,8$ are  determined by imposing the eight boundary conditions in Eq.~(\ref{BC1-inf-line}-\ref{BC4-ring}) on  $|\phi_0(x)\rangle$. We note that, on a ring the dynamics of the gap between two particles has the following symmetries $\mathcal{P}_{++}(x,t)=\mathcal{P}_{--}(x,t)=\mathcal{P}_{++}(l-x,t)$, and $\mathcal{P}_{-+}(x,t)=\mathcal{P}_{+-}(l-x,t)$ at all time $t$. Using the symmetries, the analysis simplifies a lot and we finally obtain
\begin{align}
\begin{split}
 &\mathcal{P}_{++}(x) = \mathcal{P}_{--}(x)  = \frac{1}{4Z}\lbrace A_0 + A_1 \big(e^{-k_0^{(3)} x}+e^{-k_0^{(3)} (l-x)}\big)
 +~ A_2 \big(e^{-k_0^{(5)} x}+e^{-k_0^{(5)} (l-x)}\big)\rbrace, \\ 
 &\mathcal{P}_{+-}(x) = \mathcal{P}_{-+}(l-x) = 
 \frac{1}{4Z}\lbrace A_0 + B_1\big(e^{-k_0^{(3)} x}-e^{-k_0^{(3)} (l-x)}\big)
   +\big(B_2 e^{-k_0^{(5)} x} +B_3 e^{-k_0^{(5)} (l-x)}\big)\rbrace,
 \end{split}
  \label{cont-ss-pp-pm}
 \end{align}
where $k_0^{(3)}=\sqrt{\frac{\omega}{D}},~k_0^{(5)}=\frac{\sqrt{v^2+2\omega D}}{D}$ are the spatial decay rates corresponding to $\lambda_0=0$  and the coefficients $A_0,A_1,A_2,B_1,B_2,Z$  can be expressed in terms of the $\alpha_0^{(i)}$s, and thus eventually in terms of $k_0^{(i)}$s (the explicit expressions are given in Appendix \ref{ssring-appen}). The total probability density is then given by
\begin{align}
&\mathcal{P}_{ss}(x) = \mathcal{P}_{++}(x)+\mathcal{P}_{+-}(x)+\mathcal{P}_{-+}(x)+\mathcal{P}_{--}(x)\nonumber\\
  &~=\frac{1}{4Z}\lbrace 4A_0 + 2A_1\big(e^{-k_0^{(3)} x}+e^{-k_0^{(3)} (l-x)}\big)+ \big(2 A_2 + B_2 + B_3 \big) \big(e^{-k_0^{(5)} x} + e^{-k_0^{(5)} (l-x)} \big)\rbrace. \label{cont-ss-full}
 \end{align}
In the above expression we observe that the steady state distribution is sum over two types of exponential functions characterised by two length scales $1/k_0^{(3)}=\sqrt{D/\omega}$ and $1/k_0^{(5)}=D/\sqrt{v^2+2\omega D}$ which arise due the presence of the thermal noises. This is in contrast to what Slowman {\it et al.} obtained in the problem on periodic lattice without diffusion \cite{Slowman_prl2016} who  got a single exponentially decaying `bound' configuration along with a jammed state. Note that in the $D\to 0$ limit, both length scales go to zero and one has $delta$-functions at $x=0$ and $x=l$, and a flat distribution in the middle,  similar to what was obtained in the context of single RTP in a bounded domain \cite{one-RT-particle}. The theoretical distribution in Eq.~\eqref{cont-ss-full} is verified in simulation as shown in Fig.~(\ref{steady-state-fig}) for $l=200,~D=1.0,~v=1.0,\omega=0.01$, where we observe excellent agreement. The peaks at $x=0$ and $x=l$ of the steady state distribution (representing configurations in which the two particles are  next to each other) resembles  the `equilibrium' distribution of two passive particles interacting via an effective attracting potential. From the expression of the distribution, one can identify the two body effective interaction potential as $ V_{\rm{eff}}(x)= - \ln \mathcal{P}_{ss}(x)\label{continuum-effective}$, which is attractive at short distances and saturates to a constant when the gap is large. A plot of this effective potential is given in the inset of Fig. (\ref{steady-state-fig}).

Finally, we note that, owing to translation symmetry on a ring, the joint probability distribution of the positions of the two particles in the steady state can be written as, $\mathcal{P}_{ss}(x_1,x_2) = \mathcal{P}_{ss}(X,x) = \mathcal{P}_{ss}(X)\mathcal{P}_{ss}(x),~\mbox{where}~\mathcal{P}_{ss}(X)=1/l$ is the steady state distribution of the centre of mass $X$ of the particles.

\subsection{Relaxation}
\label{continuum-ring-relaxation}
\begin{figure}[h]
\begin{center}
\includegraphics[width=12cm,angle=0.0]{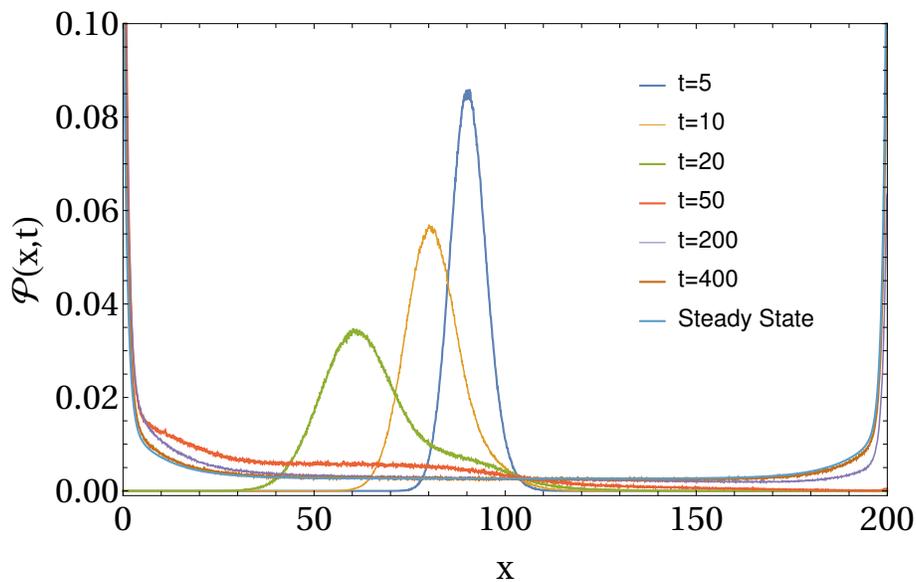}
\end{center}
\caption{Gap distribution on a ring for different times, $\mathcal{P}(x,t)$, with parameters: $l=200,~D=1.0,~v=1.0,\omega=0.1$. The initial condition is, $\mathcal{P}_{\sigma_2\sigma_1}(x,0)=\delta_{\sigma_2,1}\delta_{\sigma_1,-1}\delta(x-l/2)$.}\label{time-dep-gdf}
\end{figure}

\noindent
In this section we study the relaxation of the distribution of the gap between the two particles to the steady state discussed in the previous section.
A plot of the evolution of the probability density function $\mathcal{P}(x,t)$ for different times is shown in Fig.~(\ref{time-dep-gdf}) which describes how the distribution relaxes. 

To study this relaxation we need to consider the contribution from the eigenfunctions corresponding to non-zero eigenvalues ($\lambda_n\ne 0$) as can be seen from Eq.~\eqref{transient-distribution}. As stated earlier, the eigenvalues $\lambda_n$ and the associated eigenfunctions, more precisely the values of $\alpha_n^{(i)},~i=1,2,...,8$ can be obtained from the eight boundary conditions given in Eqs.~(\ref{BC1-inf-line}-\ref{BC4-ring}). For each $n$, we get the following eight linear homogeneous equations for the coefficients $\alpha_n^{(i)}$:
\begin{align}
\begin{split}
& \sum_{i=1}^8\alpha_n^{(i)} \bigg(\frac{v}{D}\hat{V}+k_n^{(i)}\bigg)|A_n^{(i)}\rangle=0\\
& \sum_{i=1}^8\alpha_n^{(i)} \bigg(\frac{v}{D}\hat{V}+k_n^{(i)}\bigg) e^{k_n^{(i)} l} |A_n^{(i)}\rangle=0,\\
&~\mbox{with,}~~ 
~\hat{V}=
 \begin{bmatrix}
  0 & 0 & 0 & 0\\
  0 & 1 & 0 & 0\\
  0 & 0 & -1 & 0\\
  0 & 0 & 0 & 0
 \end{bmatrix}.
 \label{Bc-lin-eq}
\end{split}
\end{align}
These equations can be recast as a matrix equation, $\mathcal{S}(k_n^{(i)})|\alpha\rangle = 0$ where $\mathcal{S}(k_n^{(i)})$ is a $8 \times 8$ matrix. The explicit expression of this matrix is provided in Appendix \ref{relaxring-appen}. Requiring non-zero solutions for the coefficients $\alpha_n^{(i)}$, demands $\text{Det}[\mathcal{S}]=0$, solving which one gets the value of $\lambda_n$ and reusing this value in Eq.~\eqref{Bc-lin-eq} and solving for  $\alpha_n^{(i)}$s one obtains $|\phi_n(x)\rangle$.

 Before proceeding further, we note that, the master equations \eqref{ME-1d-continuum} satisfied by $|\mathcal{P}\rangle$ is symmetric under certain operations, which we denote by $\mathcal{O}_s$ and $\mathcal{O}_x$: (i) $\mathcal{O}_s$: The master  equations and boundary conditions involving $P_{++}$ and $P_{--}$  in Eqs.~\eqref{ME-1d-continuum} to  (\ref{BC4-ring}) remain invariant under the transformation $+ \leftrightarrow -$ \emph{i.e.} when $P_{\sigma\sigma}\rightarrow P_{-\sigma-\sigma}$ for $\sigma = \pm 1$. (ii) $\mathcal{O}_x$: Similarly, the  master equations and the  boundary conditions remain invariant under the joint transformation $\sigma \rightarrow -\sigma$ and $x\rightarrow l-x$. Under these transformations the distributions transform as $P_{\sigma_2\sigma_1}(x)\rightarrow P_{-\sigma_2-\sigma_1}(l-x)$. Using these symmetries one can simplify the matrix equations obtained from the boundary conditions to a great extent as shown in the next section. We find that the problem of solving $\text{Det}[\mathcal{S}]=0$ of an $8 \times 8$ matrix gets reduced to setting the determinant of a smaller sized matrix to zero. In particular, it turns out that the relaxation rate {\it i.e.} $|\lambda_{1}|$ can be obtained by solving an equation involving the determinant of  a   $3\times 3$ matrix, given later in Eq.~\eqref{det-eqn-relaxation}.

Obtaining analytical expressions of the solutions of ${\rm Det}[\mathcal{S}]=0$ seems difficult. However, it is easy to explicitly check that $\lambda=-4\omega,~-2\omega,~0$ are exact solutions of $\text{Det}[\mathcal{S}]=0$ among many other solutions. We will later see in sec.~\ref{spectrum} that for large $l$ a large subset of the other solutions can be grouped around these three values with a typical level spacing of $O(l^{-2})$, as was also observed in \cite{Evans_pre-RT-spectrum} for the lattice problem without diffusion.  In the following we focus only on finding the second largest eigenvalue $\lambda_1$ for specific choices of parameters,  either numerically using Mathematica, as well as from direct numerical simulations of the equations of motion \eqref{eom_12}. In simulations, we extract the relaxation rate from the approach, with time, of the  the mean separation between the two-particles to its steady state value. At late times this approach is dominated by $\lambda_1$ and is given by $\langle x(t) \rangle - \langle x \rangle_{\rm{ss}} \propto e^{\lambda_1 t}$. 
\begin{figure}[h]
\begin{center}
\includegraphics[width=12cm,angle=0.0]{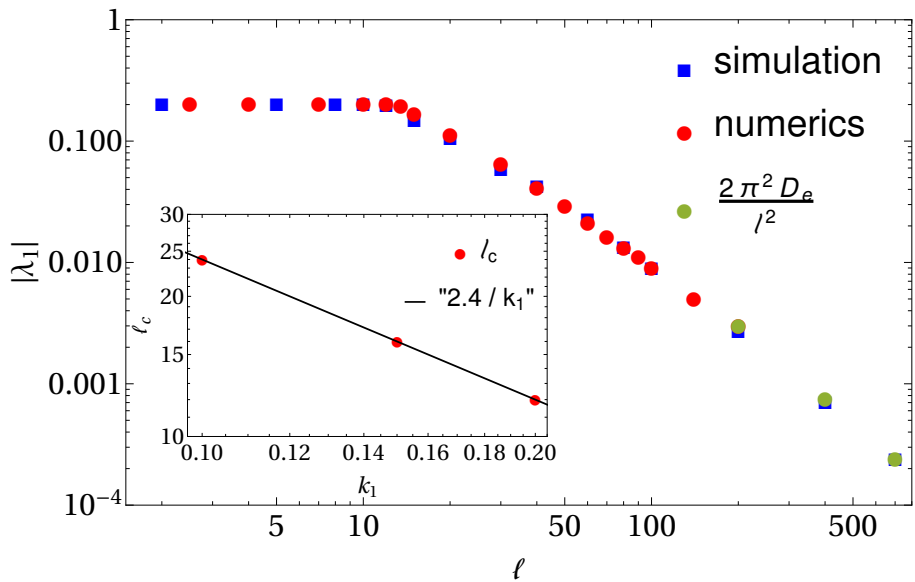}
\end{center}
\vspace{-2em}
\caption{System size dependence of the relaxation rate. For small systems the relaxation rate is a constant $2\omega$; as system increases beyond some $l_c$, $\lambda_1$ acquires a system size dependence, and at large systems the relaxation rate decreases as $1/l^{-2}$. The theoretical points are obtained from solving the determinant equation (\ref{det-eqn-relaxation}) in mathematica for small and moderate system sizes. The values for large systems are obtained by analytically solving the determinant in the large $l$ limit that gives $\lambda_1=-2\pi^2 D_e/l^2$. Here, $D=1.0,v=1.0,\omega=0.1,D_e=D+\frac{v^2}{2\omega}=6.0$. As discussed in the text, one expects $l_c \propto \sqrt{\frac{D}{\omega}}=\frac{1}{k_1}$. In the inset the dependence of $l_c$ on $k_1$ is shown for different parameters, where $l_c$ is evaluated in simulation.}
\label{relaxation-vs-l}
\end{figure}

We find that for small system size $l$, $\lambda_1$  remains at the constant value $-2w$. As $l$ is increased beyond some value $l_c$, it starts decreasing with $l$ and for very large $l (\gg l_c)$ we find $\lambda_1\propto l^{-2}$. This behaviour is demonstrated in Fig.~(\ref{relaxation-vs-l}) where we compare our theoretical result (filled discs) obtained by solving Eq.~\eqref{det-eqn-relaxation} numerically with the simulation results (filled squares). 
Heuristically, this behaviour of $\lambda_1$ with increasing  $l$ can be understood as follows. Note that for a given ring of size $l$, there are two typical time scales. One is the time scale $\tau_t \sim (2 \omega)^{-1}$, over which the spin (or the orientation) degrees of freedom relax and the other time scale is $\tau_d \sim l^2/(2D)$ over which the position degrees of freedom relax through diffusion. As a result, for small $l$, $\tau_d \ll \tau_t$,  the position variables relax fast but the spin variables do not. Hence for small $l$, relaxation is given by $\tau_r \equiv -(\lambda_1)^{-1} = \tau_t=(2\omega)^{-1}$. As $l$ increases, $\tau_d$ also increases and at $l_c\sim \sqrt{D/\omega}$ the two time scales become equal. Upon increasing $l$ further,  $\tau_d$ becomes larger than $\tau_t$ and the magnitude of $\lambda_1$ starts decreasing. When the system size becomes very large the relaxation time also becomes quite large and within this time the system experiences a large number of tumble events, the collective contribution of which can be thought of an additional white noise  [independent of the existing thermal noise $\eta_i(t)$] acting on a passive particle with an effective diffusion constant $\frac{v^2}{2\omega}$. As a result, relaxation of the gap can be effectively obtained from the relaxation of a passive Brownian particle inside a box of size $l$ with diffusion constant $D_e=D+\frac{v^2}{2\omega}$  (see Appendix \ref{relaxring-appen} for details). The corresponding relaxation time is given by $\tau_r = l^2/(2 \pi^2 D_e)$ suggesting a power law decay of the relaxation rate at large $l$ which is observed in the simulation also, as shown in Fig.~(\ref{relaxation-vs-l}). In the inset of Fig.~(\ref{relaxation-vs-l}), the simulation results for the dependence of $l_c$ on system parameters is shown. 
We note that a similar dynamical transition of the relaxation time was observed in \cite{Evans_pre-RT-spectrum} for one and two RTPs on a periodic lattice, where a similar physical explanation was proposed.

\begin{figure}[t]
\begin{center}
\includegraphics[width=12cm,angle=0.0]{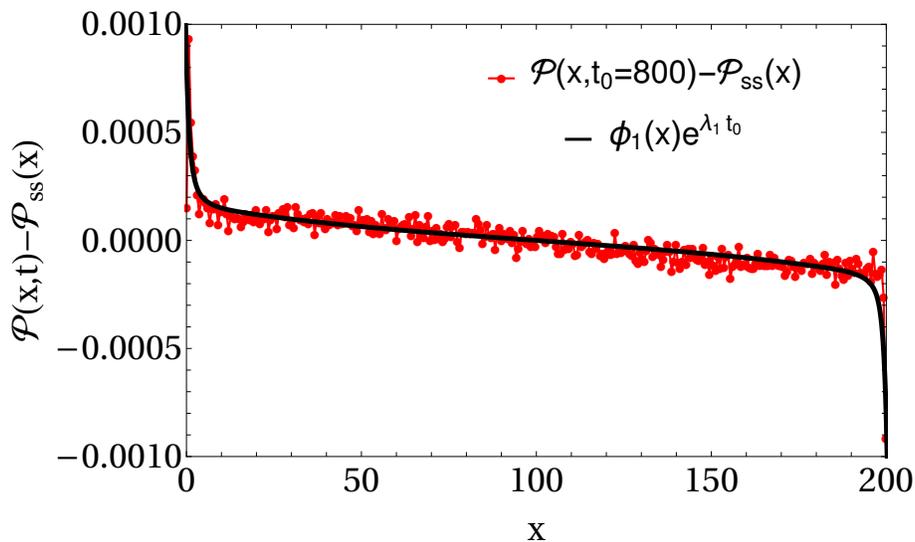}
\end{center}
\vspace{-2em}
\caption{For parameters, $v=1,D=1,\omega=0.1,l=200$, the theoretically obtained relaxation rate $|\lambda_1|\approx 0.003$ and the corresponding eigenfunction $\phi_1(x)e^{\lambda_1 t_0}$ is compared with simulation at a time $t_0=800\gg \omega^{-1}$.}\label{fig-relaxation}
\end{figure}

{An alternate way to numerically verify the relaxation time in addition to the corresponding eigenfunction, is to look at the late time convergence of distribution $\mathcal{P}(x,t)$ to the steady state distribution $\mathcal{P}_{\rm ss}(x)$ itself, where one has 
$\mathcal{P}(x,t)-\mathcal{P}_{\rm ss}(x)\approx c_1 \phi_1(x) e^{\lambda_1 t}$.
In Fig.~(\ref{fig-relaxation}) we plot $\mathcal{P}(x,t_0)-\mathcal{P}_{\rm ss}(x)$ computed numerically at a large time $t_0$ as a function of $x$ which is compared with $c_1 \phi_1(x)e^{\lambda_1t_0}$ {\it vs.} $x$ where the eigenvalue $\lambda_1$ and the corresponding eigenfunction $\phi_1(x)$ are computed theoretically (see section \ref{spectrum} and \ref{relaxring-appen}). The constant $c_1$ can in principle be obtained from initial configuration. We observe excellent agreement.}

\subsection{Eigenvalue spectrum}
\label{spectrum}
In the previous sections we have mainly  discussed the two largest eigenvalues $0$ and $\lambda_1$ which describe respectively the steady state and the approach to it. We expect that the operator $\mathcal{L}$ describing evolution inside a bounded domain of size $l$ has a  spectrum of (discrete) eigenvalues. In this section we study this spectrum which, in principle, can be obtained by solving $\text{Det}[\mathcal{S}]=0$ for $\lambda$. This task seems to be quite hard. However, as mentioned in the previous section, one can exploit the symmetries under operations $\mathcal{O}_s$ and $\mathcal{O}_x$ to simplify the problem considerably.

From the definitions, it is clear that both the symmetry operations satisfy $\mathcal{O}_s^2=\mathcal{O}_x^2={\bf I}$, where ${\bf I}$ is the Identity operator. It can be shown that the evolution operator $\mathcal{L}$  and the symmetry operators $\mathcal{O}_s$ and $\mathcal{O}_x$ constitute a set of commuting operators. Therefore the (right) eigenfunctions $|\phi_n\rangle$'s of $\mathcal{L}$ can be made simultaneous eigenfunctions of $\mathcal{O}_s$ and $\mathcal{O}_x$ as well with corresponding eigenvalues $o_s=\pm 1$ and $o_x=\pm 1$, respectively. Thus the four choices of $(o_s,o_x)$ divide the whole spectrum into four symmetry sectors --- even-even, even-odd, odd-even, odd-odd --- characterised by the eigenfunctions obeying these symmetries. Below we discuss these sectors separately in detail. For this we rewrite the eigenfunctions given in Eq.~\eqref{eigenvec} along with Eq.~\eqref{evectors}, explicitly as
\begin{eqnarray} \label{eigenvec_1}
|\phi(x) \rangle &=& \big(\alpha^{(1)} e^{k^{(1)} x}+\alpha^{(2)} e^{-k^{(1)} x}\big)
\begin{bmatrix}
   1 \\
   0 \\
   0 \\
   -1
\end{bmatrix}
 ~+~\sum_{i=3}^8 \alpha^{(i)} e^{k^{(i)} x} 
  \begin{bmatrix}
   1 \\
   a_2(k^{(i)}) \\
   a_3(k^{(i)})\\
   1
  \end{bmatrix}.
\end{eqnarray}

\subsubsection{Sectors with $o_s=-1$ and $o_x=\pm1$}
~\\
In this case the eigenfunctions should satisfy $\mathcal{O}_s|\phi\rangle=-|\phi\rangle$. 
Since, by definition
$\mathcal{O}_s (\phi^{++},\phi^{+-},\phi^{-+},\phi^{--})^T = (\phi^{--},\phi^{+-},\phi^{-+},\phi^{++})^T$, this implies that for $o_s=-1$ we have $ \alpha^{(i)}=0 ~ i=3,4,...,8$.
We still need to find $\alpha^{(1)}$ and $\alpha^{(2)}$. The boundary conditions  $\frac{\partial \phi^{\sigma \sigma}}{\partial x}|_{x=0}=0$ for $\sigma=\pm1$ at $x=0$ imply $\alpha_1=\alpha_2$ whereas the boundary condition $\frac{\partial \phi^{\sigma \sigma}}{\partial x}|_{x=l}=0$ for $\sigma=\pm1$ at $x=l$ imply $e^{2 k^{(1)} l}=1$ \emph{i.e.} $e^{k^{(1)} l}=\pm1$. To choose between these two signs, we impose that these eigenfunctions are also simultaneous eigenfunctions of $\mathcal{O}_x$. We have two cases ---
\noindent
\paragraph*{Case(i)}: If the eigenvalue of $\mathcal{O}_x$ is $o_x=-1$, then the eigenfunctions are `anti-symmetric' {\it i.e.} $\phi^{\sigma_1,\sigma_2}(x)=-\phi^{-\sigma_1,-\sigma_2}(l-x)$ for $\sigma_{1,2}=\pm1$. This implies $e^{k^{(1)} l}=1$ whose solutions are 
\begin{align}
\lambda_{n}=-2\omega-\frac{2D\pi^2 (2n)^2}{l^2}~\text{for}~n=0,1,2,... \label{sector11}
\end{align}
\paragraph*{Case(ii)}: On the other hand if the eigenvalue of $\mathcal{O}_x$ is $o_x=1$, then the eigenfunctions are `symmetric' {\it i.e.} $\phi^{\sigma_1,\sigma_2}(x)=\phi^{-\sigma_1,-\sigma_2}(l-x)$ for $\sigma_{1,2}=\pm1$ which implies $e^{k^{(1)} l}=-1$. Solutions of this equation are 
\begin{align}
\lambda_{n}=-2\omega-\frac{2D\pi^2 (2n+1)^2}{l^2}~\text{for}~n=0,1,2,... \label{sector12}
\end{align}
These solutions do not contribute to the full gap distribution $\mathcal{P}(x,t)$ and have no role in its relaxation.

\subsubsection{Sectors with $o_s=1$ and $o_x=\pm1$}
~\\
Following the same procedure, we now observe that in this case we have $\alpha^{(1)}=\alpha^{(2)}=0$ because the eigenfunctions should satisfy $\mathcal{O}_s|\phi\rangle=|\phi\rangle$. To find the remaining $\alpha^{(i)}$s we impose the condition that these eigenfunction should also be simultaneous eigenfunctions of $\mathcal{O}_x$ with eigenvalues $o_x=\pm1$. The two cases give ---
\noindent \paragraph*{Case(i)}:
For the case $o_x=-1$ we have $\phi^{\sigma_2,\sigma_1}(x)=-\phi^{-\sigma_2,-\sigma_1}(l-x)$, which implies,
\begin{eqnarray}
 \alpha^{(3)}=-\alpha^{(4)} e^{-k_a l}, 
 \alpha^{(5)}=-\alpha^{(6)} e^{-k_b l}, 
 \alpha^{(7)}=-\alpha^{(8)} e^{-k_c l}.~~~~~ \label{asym_coeffs}
\end{eqnarray}
Inserting these relations in equation~\eqref{eigenvec_1} and using the resulting eigenvector in the first three boundary conditions in Eqs.~\eqref{BC1-inf-line}-\eqref{BC4-ring} (the symmetry ensures that other boundary conditions are automatically satisfied), we get three linear homogeneous equations for the three unknowns $ \alpha^{(3)},~ \alpha^{(5)}$ and $ \alpha^{(7)}$. To find the non-trivial solutions we need to equate the determinant of the corresponding matrix to zero which provides us the following equation
~
\begin{equation}
\mbox{Det}
\begin{bmatrix}
k_a\big(1+e^{-k_a l}\big) & k_b\big(1+e^{-k_b l}\big) & k_c\big(1+e^{-k_cl}\big)\\
S_3(k_a)+S_2(k_a)e^{-k_a}l & S_3(k_b)+S_2(k_b)e^{-k_b l} & S_3(k_c)+S_2(k_c)e^{-k_c l}\\
S_2(k_a)+S_3(k_a)e^{-k_a l} & S_2(k_b)+S_3(k_b)e^{-k_b l} & S_2(k_c)+S_3(k_c)e^{-k_c l}
 \end{bmatrix}
 =0,\label{det-eqn-relaxation}
\end{equation}
where, $S_2(k)=(k+v/D)a_2(k)~\mbox{and}~S_3(k)=(k-v/D)a_3(k)$.

\noindent \paragraph*{Case (ii)}:
For the case $o_x=1$ we have $\phi^{\sigma_2,\sigma_1}(x)=\phi^{-\sigma_2,-\sigma_1}(l-x)$, which implies
\begin{eqnarray}
 \alpha^{(3)}=\alpha^{(4)} e^{-k_a l},~ 
 \alpha^{(5)}=\alpha^{(6)} e^{-k_b l},~  
 \alpha^{(7)}=\alpha^{(8)} e^{-k_c l}.~~\label{sym_coeffs} 
\end{eqnarray}
In a similar manner, inserting these relations in equation~\eqref{eigenvec_1} and using the resulting eigenvector in the first three boundary conditions in Eqs.~\eqref{BC1-inf-line}-\eqref{BC4-ring} and using the condition of having non-trivial solutions for the $\alpha$'s, we end up with the following determinant equation:
~
\begin{equation}
\mbox{Det}
\begin{bmatrix}
k_a\big(1-e^{-k_al}\big) & k_b\big(1-e^{-k_bl}\big) & k_c\big(1-e^{-k_cl}\big)\\
S_3(k_a)-S_2(k_a)e^{-k_a}l & S_3(k_b)-S_2(k_b)e^{-k_bl} & S_3(k_c)-S_2(k_c)e^{-k_cl}\\
S_2(k_a)-S_3(k_a)e^{-k_al} & S_2(k_b)-S_3(k_b)e^{-k_bl} & S_2(k_c)-S_3(k_c)e^{-k_cl}
 \end{bmatrix}
 =0.\label{det-eqn-sym}
\end{equation}
~
Solving the equations in \eqref{det-eqn-relaxation} and \eqref{det-eqn-sym} for $\lambda$, one can in principle find the rest of all possible values of $\lambda_n$'s. The associated eigenfunction is obtained by putting back this $\lambda$ value in the boundary conditions and solving for the three independent coefficients ($\alpha$'s), both `symmetric' and `anti-symmetric'.  It turns out that even these simplified zero-determinant equations \eqref{det-eqn-relaxation} and \eqref{det-eqn-sym} still give rise to very complicated transcendental equations which are hard to solve. Numerically they can be solved only for small system sizes. However, it seems possible to make some progress analytically in the large $l$ limit which we present in the next section.

\subsubsection{Structure of spectrum}
\label{Band}
~\\
\noindent
As we have seen in the previous section, the eigenvalue spectrum can be classified into four sectors based on symmetry considerations. In general the spectrum is complex and determining their structure is difficult.
However, in the large $l$ limit, it is possible to analytically compute the spectrum around the values $\lambda^*= 0,~-2\omega$ and $-4 \omega$ as exact solutions of the equation $\text{Det}[\mathcal{S}]=0$. 

Note that the complete contribution from the sector $(o_s=-1,~o_x=\pm1)$ has already been computed and given explicitly in Eqs.~\eqref{sector11} and \eqref{sector12}. Contribution from the other sector $(o_s=1,~o_x=\pm1)$ to the above mentioned three groups can be computed 
perturbatively in the $l \to \infty$ limit. To do that we consider small deviation around $\lambda^*$ {\it i.e.} $\lambda = \lambda^* + \Delta$ where $\Delta$ is very small for large $l$. Consequently from Eqs.~\eqref{ks}  we have $k^{(i)}=k^{(i)*}+\delta k^{(i)}$ for $i=1,2,...,8$ where $\delta k^{(i)}$s (determined by $\Delta$) are very small and  $k^{(i)*}$s associated to $\lambda^*$ are the eight spatial decay modes associated to $\lambda*$. Inserting these $k^{(i)}$s in Eqs.~\eqref{det-eqn-relaxation} and \eqref{det-eqn-sym} we solve for $\delta k^{(i)}$s in the leading order for large $l$. Below we present these solutions around the three values of $\lambda^*$ separately.
\noindent
\paragraph*{ Eigenvalues near $\lambda^*=0$}: 
In this case, putting $\lambda = \Delta$ in Eqs.~\eqref{ks} one finds $k_a^2=\frac{\omega}{D} + \frac{\Delta}{2D}$, $k_b^2 \simeq \frac{v^2+2\omega D}{D^2}+\frac{\Delta}{D} \frac{v^2+\omega D}{v^2+2\omega D}+O(\Delta^2)$ and $k_c^2 \simeq \frac{\Delta}{D} \frac{\omega D}{v^2+2\omega D}+O(\Delta^2)$ for small $\Delta$. Inserting these expressions in Eq.~\eqref{det-eqn-relaxation} and performing simplifications, we find that the determinant equation becomes
\begin{equation}
 \big(1+e^{-\sqrt{\frac{\Delta}{2D_e}}l}\big) + O(\sqrt{\Delta})=0.
\end{equation}

\noindent
This equation, in the leading order, implies $e^{-l\sqrt{\frac{\Delta}{2D_e}}}=-1$,  solving which one gets $\lambda_n \approx-\frac{2\pi^2 D_e (2n-1)^2}{l^2},~n\ge 1$. Using these solutions in the matrix equation, obtained from the boundary conditions, one can determine the corresponding eigenfunctions to leading order in $\Delta$. Note that these eigenstates  belong to the even-odd sector corresponding to $o_s=1,~o_x=-1$. To get the eigenstates in the even-even sector ($o_s=1,~o_x=1$), we insert the above approximate expressions of $k_a,~k_b$ and $k_c$ in  \eqref{det-eqn-sym} and follow the same procedure which finally provides 
$\lambda_n \approx -\frac{2\pi^2 D_e (2n)^2}{l^2},~n\ge 1$. Note that the above approximate solutions (in both the sectors) are valid as long as $|\lambda_n| \ll 2\omega$ {\it i.e.} for $n \ll \frac{l}{2\pi} \sqrt{\frac{\omega}{D_e}}=\sqrt{\frac{\tau_r}{2\tau_t}}$. One can improve upon the above approximate expressions of the eigenvalues by taking the terms higher order in $\Delta$ into account carefully.

\noindent
\paragraph*{Eigenvalues near $\lambda^*=-2\omega$}: To find the eigenvalues near $\lambda^*=-2\omega$ we insert $\lambda=-2\omega+\Delta$ in Eqs.~\eqref{ks} and expanding the right hand sides for $\Delta$ small we get ---
$k_a = \sqrt{\Delta/2 D}$, $k_{b}\approx \frac{\sqrt{v^2+ \sqrt{v^4+4D^2\omega^2}}}{D\sqrt{2}}+O(\Delta)$ and $k_c\approx \frac{\sqrt{v^2- \sqrt{v^4+4D^2\omega^2}}}{D\sqrt{2}}+O(\Delta)$. Note that the leading order term of $k_c$ is purely imaginary.
Substituting these expressions for the $k$'s in the determinant equations (\ref{det-eqn-relaxation}) and (\ref{det-eqn-sym}) and simplifying, we get,
\begin{equation}
(1 \mp e^{-\sqrt{\frac{\Delta}{2D}}l}) + O(\sqrt{\Delta})=0.
\end{equation}
\noindent
The equations with $-$ and $+$ signs come from Eqs.~(\ref{det-eqn-relaxation}) and (\ref{det-eqn-sym}), respectively.
Solving these equations, one obtains $\lambda_n\approx-2\omega-\frac{2\pi^2 D (2n-1)^2}{l^2}$ in the even-odd sector, and $\lambda_{n}\approx-2\omega-\frac{2\pi^2 D (2n)^2}{l^2}$ in the even-even sector with $n$ being nonzero positive integers constrained by the requirement, $|\Delta_n|<<2\omega$.

\paragraph*{Eigenvalues near $\lambda^*=-4\omega$}: To find eigenvalues near $\lambda^*=-4\omega$, we follow the same procedure as done in the previous two cases. We find that for $v^2 \neq 2 \omega D$, $\lambda_n=-4\omega +\frac{\pi^2(v^2-2\omega D)n^2}{\omega l^2}$ for $n=1,2,...$ where the odd values of $n$ fall in the even-odd sector whereas the even values fall in the even-even sector. The case $v^2=2\omega D$ is subtle and has to be analyzed carefully because in this case inserting  $\lambda=-4\omega$ in Eqs.~\eqref{ks} one gets both $k_b=0$ and $k_c=0$. Consequently, for $\lambda=-4 \omega + \Delta$ one has $k_a=i\sqrt{\frac{\omega}{D}}+O(\Delta)$, $k_b\approx\frac{(\omega \Delta)^{1/4}}{\sqrt{D}}$ and $k_c\approx i\frac{(\omega \Delta)^{1/4}}{\sqrt{D}}$. Inserting these expressions in Eqs.~(\ref{det-eqn-relaxation}) and (\ref{det-eqn-sym}) and after performing some simplifications, one gets 
\begin{align}
\tan(T/2)\pm \tanh(T/2)+O(\Delta^{1/4})=0,
\end{align}
with $T=l\frac{(\omega \Delta)^{1/4}}{\sqrt{D}}$, solving which provides $\lambda_n \approx -4 \omega +\frac{\pi^4 D^2 (4n\mp1)^4}{16~\omega l^4}$ with $n=1,2,3...$. The signs $-$ and $+$ signs correspond to the even-odd and even-even sectors, respectively.

\section{Solution on infinite line}\label{infinite_line}
\begin{figure}[t]
\begin{center}
\includegraphics[width=12cm,angle=0.0]{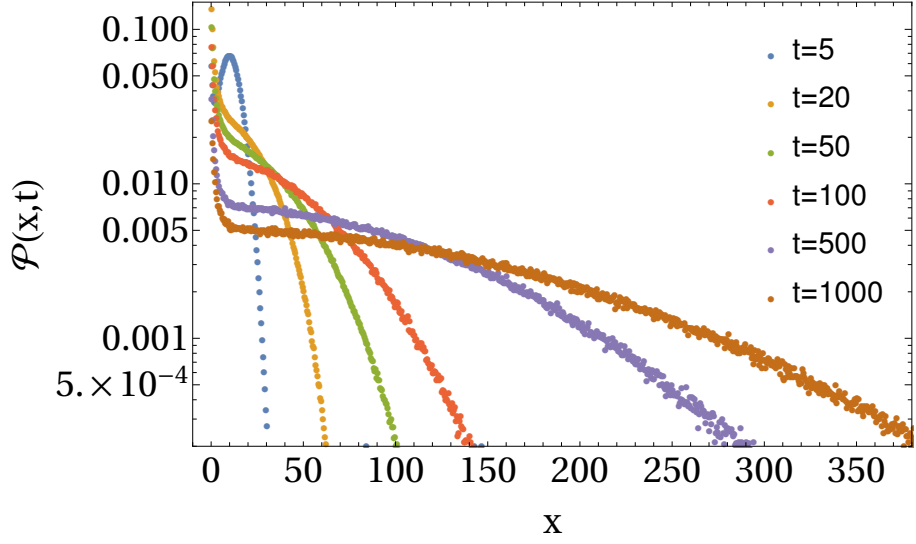}
\end{center}
\vspace{-2em}
\caption{Gap distribution for $\omega=0.1$ at different times. In the inset the distribution is plotted in semi-log scale.}\label{infline-alltime}
\end{figure}

\begin{figure}[h]
\begin{center}
\includegraphics[width=12cm,angle=0.0]{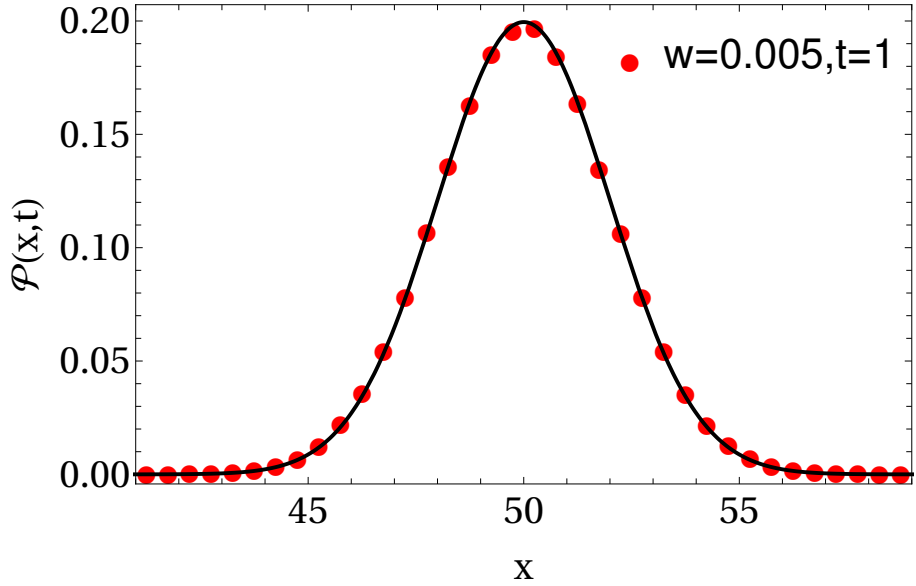}
\end{center}
\vspace{-2em}
\caption{Gap distribution for the RTPs at time $t\ll \tau_t,\tau_0$ with initial condition, $\mathcal{P}_{\sigma_2\sigma_1}(x,0)=\delta_{\sigma_21}\delta_{\sigma_11}\delta(x-x_0)$. The curve corresponds to the expression for Brownian walker of diffusivity $2D$ with a reflecting wall at origin.}\label{short-time-inf-line}
\end{figure}

In this section we study the distribution of the gap between the two RTPs moving on the infinite line, starting from a 
separation $x_0$. As the two particles evolve the distribution of the gaps starts broadening, as can be seen in the Fig.~(\ref{infline-alltime}) where we plot this distribution at different times. On the infinite line the gap distribution does not reach a steady state and keeps on evolving.  In this section we study this evolution. 
Notice that there are two typical time scales in the full problem, viz. $\tau_t=(2\omega)^{-1}$, and $\tau_0=\frac{x_0^2}{D_e}$. At very short times $t\ll \tau_t$, the particles are unlikely to undergo any tumble event and  they are performing two independent biased diffusion. As a result at such time scales, the gap distribution is given by the distribution of the separation between a pair of biased passive Brownian particles each diffusing with diffusion constant $D$ and it can be easily obtained by considering an effective Brownian particle of diffusion constant $2D$ in presence of a reflecting wall at the origin. This feature is verified numerically in  Fig.~(\ref{short-time-inf-line}).
After the time becomes comparable to $\tau_t$, the particles start changing their direction of velocity and the effect of activity starts appearing in the distribution. In order to find the distribution in such time scales one needs to solve the  
master equations \eqref{ME-1d-continuum} with the corresponding boundary conditions \eqref{BC1-inf-line}-\eqref{BC4-inf-line}  and \eqref{BC-infinity}. It, however, seems difficult to solve these equations for arbitrary time. We instead try to solve them in the large time limit and see how the presence of activity affects the distribution.

At very large times {\it i.e.} $t\gg \tau_t,\tau_0$, the particles undergo many tumble events. Therefore for large $x~\mbox{and}~t$ the particles behave like independent Brownian particles but now with effective diffusion constant  $D_e=D+\frac{v^2}{2\omega}$ as explained previously [ also see \cite{Kanaya-ABP}], and consequently the corresponding two-particle gap distribution would be that of an effective Brownian walker with diffusion constant $2D_e$ moving in a semi-infinite line with a reflecting wall at $x=0$. In this limit, it is easy to see that the gap distribution should have the following diffusive  scaling 
\begin{equation}
\mathcal{P}(x,t) \xrightarrow{x\rightarrow \infty, t\rightarrow \infty} \frac{1}{\sqrt{t}}\Phi\left(\frac{x}{\sqrt{t}}\right), \text{where,}~~
\Phi(u)=\frac{e^{-\frac{u^2}{8 D_e}}}{\sqrt{2\pi D_e}}\label{inf-line-scaling}.
\end{equation}
\begin{figure}[h]
\begin{center}
\includegraphics[width=12cm,angle=0.0]{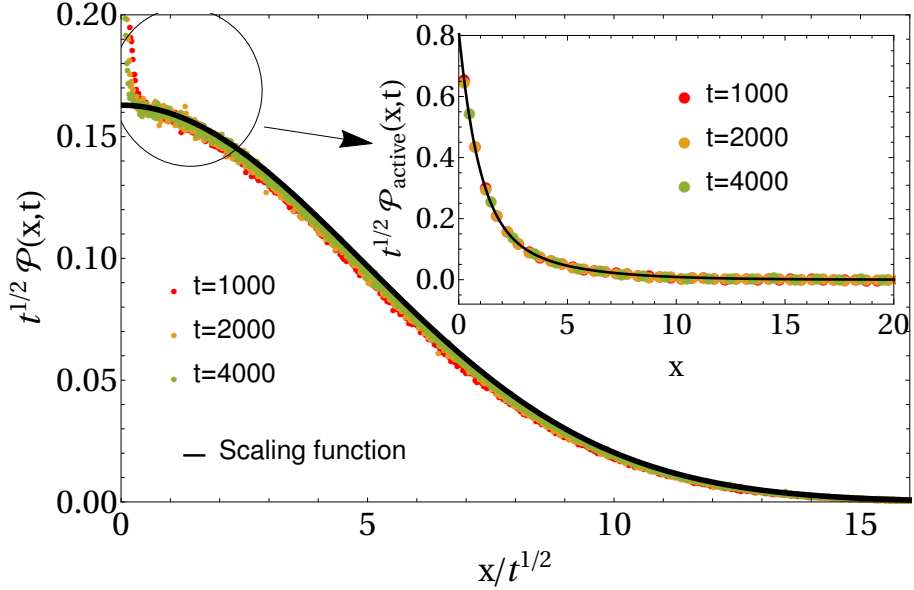}
\end{center}
\vspace{-2em}
\caption{Scaling collapse for the gap distribution on infinite line in large times. The parameters are, $D=1.0,v=1.0,\omega=0.1,x_0=1.0$. The inset shows numerical verification of the scaling behaviour of the distribution for small values of gaps as predicted by  Eq.~\eqref{smallgap}.}
\label{collapse-infline}
\end{figure}
We observe the existence of this scaling behaviour numerically from the collapse of data points collected at different times in Fig.~(\ref{collapse-infline}). It is important to note that this scaling form is valid only over a range of order $O(\sqrt{t})$ {\it i.e.} $(x-x_0) \sim \sqrt{2D_et}$. Therefore the distribution near $x =0$ cannot be described correctly by the above scaling form, because we expect to see the effect of activity in this region and one cannot approximate the dynamics of the two particles by independent effective Brownian dynamics. In fact, we perform a careful analysis of the solution of the master equations and find that for large $t$ (see Appendix \ref{Appendix_B} for details) the distribution is actually given by 
\begin{equation}
\lim_{t\rightarrow\infty}\mathcal{P}(x,t) \sim ~C~ \frac{e^{-\frac{x_0^2}{8 D_e t}}}{\sqrt{2\pi t}} \bigg( \frac{\sqrt{w}\sqrt{v^2+2wD}}{\sqrt{D}} e^{-k_0^{(3)} x} + \frac{v^2}{D} e^{-k_0^{(5)} x}\bigg) + \frac{1}{\sqrt{8\pi D_e t}} \big(e^{-\frac{(x+x_0)^2}{8 D_et}}+e^{-\frac{(x-x_0)^2}{8 D_e t}}\big), \label{inflinesoln}
\end{equation}
where $k_0^{(3)}=\sqrt{w/D}$, $k_0^{(5)}=\frac{\sqrt{v^2+2wD}}{D}$, $D_e=D+\frac{v^2}{2w}$, and 
$C=\frac{v^2\sqrt{D_e}}{(v^2+2wD)\big(v^2+\sqrt{wD(v^2+2wD)}\big)}$. 

From this expression one can clearly distinguish the two contributions in the distribution, one coming from the effective passive diffusion $\mathcal{P}_{diff}(x,t)=\frac{1}{\sqrt{8\pi D_e t}} \big(e^{-\frac{(x+x_0)^2}{8 D_et}}+e^{-\frac{(x-x_0)^2}{8 D_e t}}\big)$ as discussed before and the other is the `active' contribution,
\begin{equation}
\mathcal{P}_{active}(x,t)=\mathcal{P}(x,t)-\mathcal{P}_{diff}
= C \bigg( \frac{\sqrt{w}\sqrt{v^2+2wD}}{\sqrt{D}} e^{-k_0^{(3)} x} + \frac{v^2}{D} e^{-k_0^{(5)} x}\bigg) \frac{e^{-\frac{x_0^2}{8 D_e t}}}{\sqrt{2\pi t}}. ~~~\label{smallgap}
\end{equation}
Note from the above equation that, the gap distribution increases exponentially as one decreases gap towards zero.
In order to verify the `active' contribution numerically, we plot the full distribution obtained numerically after subtracting the diffusive contribution $\mathcal{P}_{diff}$  in the inset of Fig.~ (\ref{collapse-infline}) where we observe excellent agreement with the theoretical result in Eq. \eqref{smallgap} (solid lines). 

This interestingly implies that even on the infinite line the particles show considerable affinity at large times as manifested by large values for the distribution at small gap values. As expected, in the limit $v \rightarrow 0$ the result for interacting passive Brownian particles is retrieved. Finally, we mention that it is also possible to extract the behaviour in Eq.~\eqref{inflinesoln} from the time dependent results obtained in  the ring geometry, by taking $l\rightarrow \infty$ followed by $t\rightarrow \infty$ (details are discussed in Appendix \ref{Appendix_B}).

The effect of activity for interacting particles become most prominent and singular in the absence of diffusion. When external noise vanishes, the particles once collide (which can happen only if $\sigma_2=1,~\sigma_1=-1$) stick together with the inter-particle gap becoming zero. Only when both of them tumbles, the gap again starts increasing. Therefore one expects that the particles would spend a significant fraction of time in such zero-gap configurations which are called the `jammed configurations'. 

\begin{figure}[h]
\begin{center}
\includegraphics[width=12cm,angle=0.0]{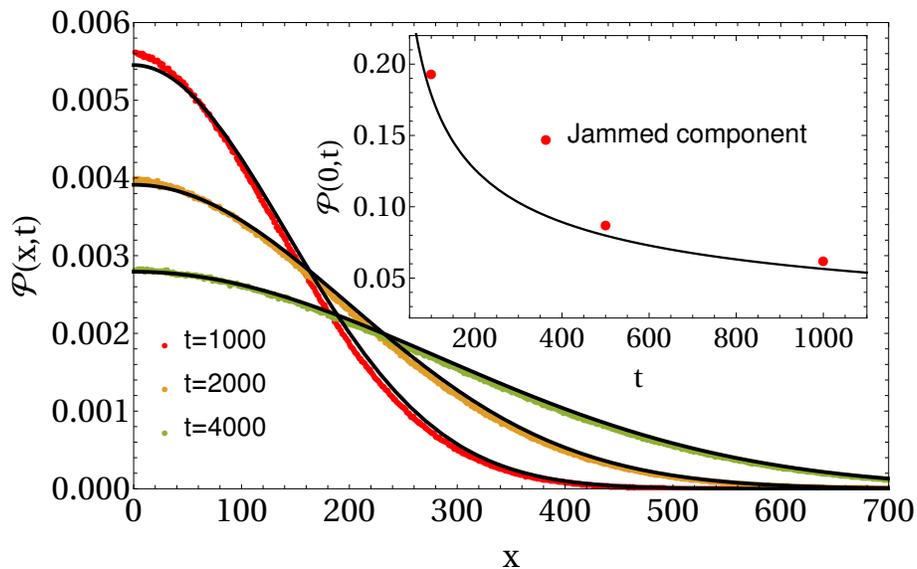}
\end{center}
\vspace{-2em}
\caption{Gap distribution for $D=0,v=1.0,w=0.1,x_0=1.0$ with $t$ very large. In the main plot the distribution except the `jammed component' is shown. The strength of the jammed component $\mathcal{P}(0,t)$ is plotted along with the term containing $\delta(x)$ in Eq.~\eqref{D0_infline} in the inset, where $\delta(x)$ is approximated by $1/{\Delta x}$ (bin size) in the simulation.}\label{no-noise}
\end{figure}

In order to see these `jammed configurations' more explicitly, we take the $D\rightarrow 0$ limit for which the equation \eqref{inflinesoln} takes the following form,
\begin{equation}
 \lim_{D\rightarrow 0} \mathcal{P}(x,t)\sim \frac{e^{-\frac{x_0^2}{8D_e t}}}{\sqrt{w\pi t}}\delta(x)+\frac{1}{\sqrt{8\pi D_e t}}
 \big(e^{-\frac{(x+x_0)^2}{8D_e t}} + e^{-\frac{(x-x_0)^2}{8D_e t}} \big),\label{D0_infline}
\end{equation}
where we have used  $\lim_{a\rightarrow \infty} a e^{-a x} = \delta(x)$ as $x\ge 0$.
Thus, in the absence of external noise, the large time behaviour is that of passive particles with an effective diffusivity $D_e=\frac{v^2}{2\omega}$, and in addition a `jammed' component (denoted by the  term containing $\delta(x)$) indeed appears with strength proportional to $1/\sqrt{t}$ for large $t$. 
Simulation results for the gap distribution for $D=0$ and the magnitude of the peak at $x=0$ are shown in
Fig.~(\ref{no-noise}).
\vspace{-2em}

\section{Conclusion}\label{summary}
In this paper we  studied the distribution of the separation between two run and tumble particles interacting via hardcore exclusion in the presence of additional thermal noise. The competition between the activity and thermal noise gives rise to a rich spatial structure  in the steady state gap distribution on a ring geometry, that is characterised by two exponentially decaying bound components. Such a steady state can equivalently be described  in terms of passive Brownian particles, again with hardcore exclusion and with an effective attractive interaction.
As the diffusivity associated to the thermal noise tends to zero, the exponentially decaying bound states become delta-function peaks, characterising the `jammed' components in the distribution as has earlier been discussed in \cite{Slowman_prl2016}.

We have found that the relaxation of the gap distribution to the steady state shows rich and nontrivial behaviour. The relaxation rate undergoes a crossover from a system-size independent value to a size dependent form as the size is increased from a small value.  At very large values of the ring size $l$, the inverse of relaxation time behaves as  $\sim 1/l^2$. Using certain symmetries of the problem we  also studied the structure of the full eigenvalue spectrum and showed that the spectrum can be grouped into four symmetry classes. Our analysis provides us with a set of analytical solutions for the decay rates $\lambda$, some of which are exact, while  others are approximate ones that are expected to be asymptotically exact in  large $l$ limit. In particular, we find that in two of the four symmetry sectors the spectrum can be grouped around $\lambda=0,-2\omega~\mbox{and}~-4\omega$ respectively. Our results on the ring thus provide an extension of the findings of \cite{Evans_pre-RT-spectrum} to the continuum model in the presence of thermal noise.

In the infinite line the gap between the two RTPs naturally does not have any steady state.  In the large time limit we obtain an asymptotic expression of the distribution of the gap analytically. Interestingly, we find that even at very large times  the gap distribution retains a  sharp peak at small values of the gap implying very high affinity even on the infinite line, and in fact in the zero noise limit, a slowly temporally decaying `jammed' component is present in the distribution.

We characterised the effect of external noise in the statistical properties of one dimensional interacting RTPs. Our results suggest that the competition between activity and external noise could be a potential microscopic mechanism that leads to very rich spatial structure and effective interaction for such systems. How do these features translate to the interacting many particle case and whether this helps to have an analytical understanding of the phase separation as observed in \cite{Levis} would be interesting questions to explore in the future.

\paragraph*{} We acknowledge support of the Department of Atomic Energy, Government of India, under project no.12-R$\&$D-TFR-5.10-1100. AD acknowledges CEFIPRA postdoctoral fellowship hosted at ICTS-TIFR. AK acknowledges support from DST grant under project No. ECR/2017/000634. AK and AD acknowledge the support of the project 5604-2 of the Indo-French Centre for the Promotion of Advanced Research (IFCPAR). \\


\appendix
\vspace{0.5em}
\section{Gap distribution on a ring}\label{Appendix_A}

\subsection{Steady state}\label{ssring-appen}
The coefficients and the normalisation constant $Z$ appeared in equations (\ref{cont-ss-pp-pm})-(\ref{cont-ss-full}) of section \ref{continuum-ring-SS} are,
 \begin{eqnarray}
  A_1 &=& 1, 
  A_2 = - \frac{k_0^{(3)} (1-e^{-k_0^{(3)} l})}{k_0^{(5)} (1-e^{-k_0^{(5)} l})}, 
  B_1 = \frac{\sqrt{w D}}{v},\nonumber\\
  B_2 &=& \frac{k_0^{(3)} (1-e^{-k_0^{(3)} l})}{k_0^{(5)} (1-e^{-k_0^{(5)} l})}\frac{wD}{v^2+wD-vDk_0^{(5)}}, 
  B_3 = \frac{k_0^{(3)} (1-e^{-k_0^{(3)} l})}{k_0^{(5)} (1-e^{-k_0^{(5)} l})}\frac{wD}{v^2+wD+vDk_0^{(5)}}, \nonumber\\
  A_0 &=& \frac{w(1-e^{-k_0^{(3)} l})}{v k_0^{(3)}}+\frac{w D(1+e^{-k_0^{(3)} l})}{v^2} + \frac{wD}{v}\frac{k_0^{(3)} (1-e^{-k_0^{(3)} l})}{k_0^{(5)} (1-e^{-k_0^{(5)} l})}\nonumber\\
  &~& ~~~~~~\times \bigg(\frac{v+Dk_0^{(5)}}{v^2+wD+vDk_0^{(5)}}+\frac{v-Dk_0^{(5)}}{v^2+wD-vDk_0^{(5)}}e^{-k_0^{(5)} l}\bigg), \nonumber\\
  Z &=& A_0 l+\frac{A_1}{k_0^{(3)}}(1-e^{-k_0^{(3)} l})+\frac{2A_2+B_2+B_3}{2k_0^{(5)}}(1-e^{-k_0^{(5)} l}),\nonumber 
 \end{eqnarray}
where, $k_0^{(3)}=\sqrt{\frac{w}{D}},~k_0^{(5)}=\frac{\sqrt{v^2+2wD}}{D}$.
 
\subsection{Transient behaviour and relaxation}\label{relaxring-appen}
In section \ref{continuum-ring-relaxation}, we have seen that the equations for the coefficients $\alpha_n^{(i)}$ obtained from the boundary conditions in Eqs. \eqref{BC1-inf-line}-\eqref{BC4-ring} could be written in the form of a matrix equation, $\mathcal{S}(k_n^{(i)})|\alpha\rangle = 0$.
For notational simplification, in the following we shall replace $\lambda_n$ by $\lambda$ and the $k_n^{(i)}$'s by $\pm \lbrace k_a,k_b,k_c \rbrace$, where $ k_a,k_b,k_c $ are given by Eqs.~\eqref{spatial-modes} and \eqref{ks}.
Recalling $S_2(k)=a_2(k)(k+v/D)$, $S_3(k)=a_3(k)(k-v/D)$ defined after Eq.~\eqref{det-eqn-relaxation}, the matrix $\mathcal{S}$ becomes,
~
\begin{align}
\mathcal{S}=
\begin{bmatrix}
  k_a & -k_a & k_a & -k_a & k_b & -k_b & k_c & -k_c\\
 0  & 0  & S_2(k_a)  & S_2(-k_a) & S_2(k_b) & S_2(-k_b)  & S_2(k_c)  & S_2(-k_c)\\
 0  & 0  & S_3(k_a)  & S_3(-k_a) & S_3(k_b)  & S_3(-k_b)  & S_3(k_c)  & S_3(-k_c)\\
 -k_a & k_a & k_a & -k_a & k_b & -k_b & k_c & -k_c\\
 k_a e^{k_a l} & -k_a e^{-k_a l} & k_a e^{k_a l} & -k_a e^{-k_a l} & k_b e^{k_b l} & -k_be^{-k_b l} & k_c e^{k_c l} & -k_c e^{-k_c l}\\
 0 & 0  & S_2(k_a) e^{k_a l}  & S_2(-k_a)e^{-k_a l} 
 & S_2(k_b)e^{k_b l} & S_2(-k_b)e^{-k_b l} & S_2(k_c)e^{k_c l} & S_2(-k_c)e^{-k_c l} \\
0  & 0  & S_3(k_a) e^{k_a l}  & S_3(-k_a)e^{-k_a l} 
 & S_3(k_b)e^{k_b l} & S_3(-k_b)e^{-k_b l} & S_3(k_c)e^{k_c l} & S_3(-k_c)e^{-k_c l} \\
 -k_ae^{k_a l} & k_ae^{-k_a l} & k_ae^{k_a l} & -k_ae^{-k_a l} & k_b e^{k_b l} & -k_b e^{-k_b l} & k_c e^{k_c l} & -k_ce^{-k_c l}
  \end{bmatrix}
  \label{mcalS-matrix}
\end{align}

The existence of nontrivial solutions for $|\alpha \rangle$'s imply ${\rm Det}[\mathcal{S}]=0$ which upon solving gives the allowed values of $\lambda$. Using the symmetries of the problem as detailed in section \ref{spectrum}, one can easily find the  set of exact solutions given in Eqs.~\eqref{sector11} and \eqref{sector12} analytically in the odd-even and odd-odd sectors. To find the solutions in the other two sectors (viz. the even-odd and even-even sectors), one needs to solve the above determinant equal to zero equation numerically. However, it turns out that once again using the symmetries one can reduce the problem to the solutions of determinants equal to zero of matrices of smaller sizes $(3 \times 3)$ given in Eqs.~\eqref{det-eqn-relaxation} and \eqref{det-eqn-sym}.\\

\textbf{I. Relaxation rate:} We first note that, the determinant equation is satisfied if either of $k_a,~k_b~\mbox{and}~k_c$ vanishes, implying that, $\lambda=-2\omega,-4\omega,0$. $\lambda=0$ gives the steady state, and we are seeking the late time relaxation rate given by the largest nonzero decay rate $\lambda_1$. For small systems $\lambda_1=-2\omega$. Finding $\lambda_1$ for arbitrary system size $l$ amounts to solving the full determinant equation, which could not be done. However, in the $l\rightarrow \infty$ limit, the relaxation rate $\lambda_1$ becomes very small. In this limit, $k_a\approx \sqrt{\frac{w}{D}},~k_b\approx \frac{\sqrt{v^2+2wD}}{D},~k_c\approx \sqrt{\frac{\lambda}{2 D_e}}$ with $D_e=D+\frac{v^2}{2w}$, and the above determinant equation after rearrangements simplifies to,
~
\[
{\rm Det}
  \begin{bmatrix}
  0 & k_a & 0 & 0 & 0 & -k_a & 0 & 0\\
 -c_1  & 0  & -d_1  & v/D & -c_2 & 0  & -d_2  & v/D\\
 c_2  & 0  & d_2  & -v/D & c_1  & 0  & d_1  & -v/D\\
 k_a & 0 & k_b & \sqrt{\frac{\lambda}{2 D_e}} & -k_a & 0 & -k_b & -\sqrt{\frac{\lambda}{2 D_e}} \\
 0 & k_a e^{k_a l} & 0 & 0 & 0 & -k_ae^{-k_a l} & 0 & 0\\
 -c_1 e^{k_a l}  & 0  & -d_1 e^{k_b l}  & v/D e^{\sqrt{\frac{\lambda}{2 D_e}} l} 
 & -c_2 e^{-k_a l} & 0  & -d_2 e^{-k_b l}  & v/D e^{- \sqrt{\frac{\lambda}{2 D_e}} l} \\
c_2 e^{k_a l}  & 0  & d_2 e^{k_b l}  & -v/D e^{\sqrt{\frac{\lambda}{2 D_e}} l} 
 & c_1 e^{-k_a l} & 0  & d_1 e^{-k_b l}  & -v/D e^{- \sqrt{\frac{\lambda}{2 D_e}} l} \\
 k_ae^{k_a l} & 0 & k_be^{k_b l} & \sqrt{\frac{\lambda}{2 D_e}} e^{\sqrt{\frac{\lambda}{2 D_e}} l} & -k_ae^{-k_a l} & 0 & -k_be^{-k_b l} & -\sqrt{\frac{\lambda}{2 D_e}} e^{-\sqrt{\frac{\lambda}{2 D_e}} l}
  \end{bmatrix} = 0, ~~~
\]
where, $c_{1,2}=\frac{w}{v}\pm \sqrt{\frac{w}{D}},~d_{1,2}=\frac{w(v/D\pm k_b)}{w+v(v/D\pm k_b)}$. The above determinant equation after some further simplification gives, $e^{\sqrt{\frac{2\lambda}{D_e}} l} = 1$. The solution of the above equation for the largest nonzero value of $\lambda$, i.e. the relaxation rate, is, $\lambda_1=-\frac{2\pi^2 D_e}{l^2}$. The large $l$ relaxation rate observed in simulation (Fig. \ref{relaxation-vs-l}) is in good agreement to that obtained above. The corresponding eigenfunction falls in the even-odd sector discussed in section \ref{Band}.

\section{RTPs on infinite line}\label{Appendix_B}

Consider the time dependent solution for the ring with the system size tending to infinity. Then, for large times the spectrum (discussed in section \ref{Band}) around $\lambda=-2\omega$ and $-4\omega$ vanish, and the time dependent distribution in equation \eqref{transient-distribution} is approximated by, 
\begin{equation}
\lim_{t\rightarrow \infty} |P(x,t)\rangle \approx \sum_n c_n e^{\lambda_n t} \sum_{i=1}^3 \alpha_i(n) \big[ e^{-k_i(n) x}|A(k_i(n))\rangle 
 +(-1)^n e^{-k_i(n) (l-x)}|A(-k_i(n))\rangle \big],\nonumber
\end{equation}
where $k_i(n)$ stands for $k_a(\lambda_n),k_b(\lambda_n)~\mbox{and}~k_c(\lambda_n)$,~ $\lambda_n=-\frac{2\pi^2 D_e n^2}{l^2}$. For very large $l,~k_a,k_b$ are approximately equal to $k_0^{(3)},k_0^{(5)}$ respectively, $k_c\propto\sqrt{\lambda_n}\sim i\frac{n}{l}$, $\alpha$'s and $|A\rangle$'s become independent of $n$ in the leading order, and $\sum_n$ is replaced by $\int_0^{\infty} dq$. Therefore the large time distribution looks like, \begin{equation}
\lim_{t,l\rightarrow \infty} |P(x,t)\rangle \approx \int_0^{\infty} dq~ c(q) e^{-\frac{1}{2} D_e q^2 t} \big[ \alpha_1 e^{-k_a x}|A(k_a)\rangle + \alpha_2 e^{-k_b x}|A(k_b)\rangle + \alpha_3 e^{-iq x}|A(0)\rangle \big],\nonumber
\end{equation}
the terms involving $e^{-k_i(n) (l-x)}$ vanishing in the large $l$ limit. The coefficients $c(q)$ can in principle be determined from the initial condition. It is however difficult to determine them for general initial condition assuming the orthogonality of the eigenstates. For initial conditions where the particles start with fixed finite gap between them (i.e. $P(x,0)=\delta(x-x_0)$), one expects $c(q)$ to be approximately proportional to $(e^{iq x_0}+e^{-iq x_0})$ as in the case of interacting Brownian walkers on the infinite line.
To obtain rigorous analytical expression for this limit, we have solved for the large time behaviour of the gap distribution on infinite line. The results are given in section \ref{infinite_line}, and below we outline the detailed steps for the calculation.

Let us choose the initial condition as, $\mathcal{P}_{\sigma_2 \sigma_1}(x,0) = p_i\delta(x-x_0)$, where $i=1(1)4$ corresponds to four possible compositions of $\lbrace \sigma_2, \sigma_1\rbrace$, and $\sum p_i = 1$.
We can recast the Fokker-Planck equations in a more symmetric form by defining, $\mathcal{P}_0 = \mathcal{P}_{++} - \mathcal{P}_{--}, 
~ \mathcal{P}_{||} = \mathcal{P}_{++} + \mathcal{P}_{--},~ \mathcal{P}_S = \mathcal{P}_{+-} + \mathcal{P}_{-+},
~\rm(and)~ \mathcal{P}_A = \mathcal{P}_{+-} - \mathcal{P}_{-+}$. These quantities satisfy the following equations,

\begin{eqnarray}
\dot{\mathcal{P}}_0&=&2D\frac{\partial^2{\mathcal{P}}_0}{\partial x^2}-2\omega \mathcal{P}_0, \label{combination_1}\\
\dot{\mathcal{P}}_{||}&=&2D\frac{\partial^2{\mathcal{P}}_{||}}{\partial x^2} + 2\omega({\mathcal{P}}_S - {\mathcal{P}}_{||}),\\
\dot{\mathcal{P}}_{S}&=&2D\frac{\partial^2{\mathcal{P}}_{S}}{\partial x^2} + 2v\frac{\partial{\mathcal{P}}_{A}}{\partial x} + 2 \omega({\mathcal{P}}_{||} - {\mathcal{P}}_{S}),\\
\dot{\mathcal{P}}_{A}&=&2D\frac{\partial^2{\mathcal{P}}_{A}}{\partial x^2} + 2v\frac{\partial{\mathcal{P}}_{S}}{\partial x} - 2\omega {\mathcal{P}}_{A},
\label{combination_4}
\end{eqnarray}
with boundary conditions,
\begin{eqnarray}
&~& \frac{\partial {\mathcal{P}}_0}{\partial x}|_{0,t}=0, \label{BC1-combination}\\
&~& \frac{\partial {\mathcal{P}}_{||}}{\partial x}|_{0,t}=0,\\
&~& D\frac{\partial {\mathcal{P}}_S}{\partial x}|_{0,t} + v{\mathcal{P}}_A(0,t)=0,\\
&~& D\frac{\partial {\mathcal{P}}_A}{\partial x}|_{0,t} + v{\mathcal{P}}_S(0,t) =0, \label{BC4-combination}\\
&~& |\mathcal{P}(x\rightarrow \infty,t)\rangle = 0. \label{BC-infty}
\end{eqnarray}
The gap distribution is given by $\mathcal{P}(x,t)=\mathcal{P}_{||} + \mathcal{P}_S$.
The equation for $\mathcal{P}_0(x,t)$ is immediately solved,
\begin{equation}
 \mathcal{P}_0(x,t) = \frac{p_1 - p_4}{\sqrt{8\pi D t}}e^{-2wt}\big(e^{-(x+x_0)^2/8Dt} + e^{-(x-x_0)^2/8Dt} \big).\label{paralleldiff}
\end{equation}
To solve the other three quantities, we first take the Laplace transform,
${\rm LT}_t\lbrace|\mathcal{P}(x,t)\rangle \rbrace = |\tilde{\mathcal{P}}(x,s)\rangle = \int_0^\infty dt e^{-s t} |\mathcal{P}(x,t)\rangle$ to get, from the equations,
\begin{eqnarray}
- (p_1+p_4) \delta(x-x_0) &=&2D\frac{\partial^2{\tilde{P}}_{||}}{\partial x^2} -(s+2\omega){\tilde{P}}_{||} +
2\omega \tilde{P}_S, \label{temp-laplace_1}\nonumber\\
- (p_2+p_3) \delta(x-x_0) &=&2D\frac{\partial^2{\tilde{P}}_S}{\partial x^2}+2v\frac{\partial{\tilde{P}}_A}{\partial x}-(s+2\omega){\tilde{P}}_S + 2\omega{\tilde{P}}_{||},\nonumber\\
- (p_2-p_3) \delta(x-x_0) &=&2D\frac{\partial^2{\tilde{P}}_{A}}{\partial x^2} + 2v\frac{\partial{\tilde{P}}_{S}}{\partial x}-(s+2\omega){\tilde{P}}_A~\label{temp-laplace_3},\nonumber
\end{eqnarray}
 and the boundary conditions,
 \begin{eqnarray}
&~& \frac{\partial \tilde{P}_{||}}{\partial x}|_{0,s}=0, \nonumber\\
&~& D\frac{\partial \tilde{P}_S}{\partial x}|_{0,s} + v \tilde{P}_A(0,s)=0,\nonumber\\
&~& D\frac{\partial \tilde{P}_A}{\partial x}|_{0,t} + v \tilde{P}_S(0,s)=0,\nonumber\\
&~& |\tilde{P}(x\rightarrow\infty,s)\rangle = 0. \nonumber
\end{eqnarray}
Next we perform the Laplace transform of $\tilde{P}$ over $x$ as, $|Q(\mu,s)\rangle = \int_0^\infty dx e^{-\mu x} |\tilde{P}(x,s)\rangle$. The $Q(\mu,s)$'s satisfy the following algebraic equations,
\begin{eqnarray}
&&(2D\mu^2 - 2w - s)Q_{||} + 2w Q_S = 2D\mu \tilde{P}_{||}(0,s) - (p_1+p_4)e^{-\mu x_0},\nonumber\\
&& 2w Q_{||} + (2D\mu^2 - 2w - s)Q_S + 2v\mu Q_A = 2D\mu \tilde{P}_S(0,s) - (p_2+p_3)e^{-\mu x_0},\nonumber\\
&& 2v\mu Q_S + (2D\mu^2 - 2w - s)Q_A = 2D\mu \tilde{P}_A(0,s) - (p_2-p_3)e^{-\mu x_0}.\nonumber
\end{eqnarray}
Before proceeding, we note that the long time behaviour does not depend on the initial orientation. The solution for $\mathcal{P}_0(x,t)$ in equation (\ref{paralleldiff}) decreases with time as $e^{-2wt}$ implying that the $++$ and $--$ components become identical at very large times. More generally, although the internal active noise $\Sigma=\sigma_1-\sigma_2$ is a non-stationary process with correlations depending upon the exact time and initial orientation, the information about the initial orientation is lost exponentially with time, $\langle \Sigma(t)\Sigma(t')\rangle=2\big(e^{-2w|t-t'|}-\sigma_{10}\sigma_{20}e^{-2w(t+t')}\big)$, $\sigma_{10}, \sigma_{20}$ being the initial orientations of the two particles. The independence of $\mathcal{P}_{\sigma_2\sigma_1}$ from initial orientations of the particles in the long time limit is also verified using simulation. We have chosen, $p_1=1,p_2=p_3=p_4=0$, i.e. the initial condition where both the particles start moving in the positive $x$ direction at $t=0$ with a gap $x_0$. The components of $|Q(\mu,s)\rangle$ for this initial condition is obtained by solving the algebraic equations as,
\begin{eqnarray}
 Q_{||} &=& \frac{\mu \tilde{P}_{||}(0,s) - \frac{e^{-\mu x_0}}{2D} - \frac{w}{D}Q_S}{\mu^2 - \mu_0^2},\nonumber\\
  Q_A &=& \mu \frac{\tilde{P}_A(0,s) - \frac{v}{D}Q_S}{\mu^2 - \mu_0^2},\nonumber\\
 Q_S &=& -\frac{w\mu \tilde{P}_{||}(0,s)}{D(\mu^2-\mu_1^2)(\mu^2-\mu_2^2)} + \frac{\mu(\mu^2-\mu_0^2)\tilde{P}_S(0,s)}{(\mu^2-\mu_1^2)(\mu^2-\mu_2^2)}\nonumber\\
 && - \frac{v \mu^2 \tilde{P}_A(0,s)}{D(\mu^2-\mu_1^2)(\mu^2-\mu_2^2)} + \frac{w e^{-\mu x_0}}{2D^2(\mu^2-\mu_1^2)(\mu^2-\mu_2^2)},\nonumber
\end{eqnarray}
where,
\begin{eqnarray}
\mu_i^2 =
\left\{
\begin{array}{ll}
&\frac{\omega + \frac{s}{2}}{D} \cr
&\frac{2\omega D + v^2}{2D^2}+\frac{s}{2D}+\frac{\sqrt{(2\omega D+v^2)^2+2s D v^2}}{2D^2}\cr
&\frac{2\omega D + v^2}{2D^2}+\frac{s}{2D}-\frac{\sqrt{(2\omega D+v^2)^2+2s D v^2}}{2D^2}
\end{array}
\right.
\end{eqnarray}
with $i=0,1,2$.
We are interested in calculating the full gap distribution $\mathcal{P}(x,t)$ at late times. For this, we first invert $Q(\mu,s)=Q_{||}+Q_S$ in $\mu$ to get $\tilde{P}(x,s)$ in the small $s$ limit. In this limit, $\mu_0\approx k_0^{(3)}=\sqrt{\frac{w}{D}},~\mu_1\approx k_0^{(5)}=\frac{\sqrt{v^2+2wD}}{D},~\mu_2\approx \sqrt{\frac{s}{2 D_e}}$ with $D_e=D+\frac{v^2}{2w}$. Using these we obtain,
\begin{eqnarray}
&&\lim_{s\rightarrow 0} \tilde{P}(x,s)= \lim_{s\rightarrow 0} {\rm LT}_{\mu}^{-1}\lbrace Q_{||}+Q_S\rbrace\nonumber\\
 && = \tilde{P}_{||}(0,s) \bigg[\frac{v^2}{v^2+wD}\big(\cosh\mu_0x - \frac{\mu_0^2}{\mu_1^2}\cosh\mu_1x \big) + \frac{2\mu_0^2}{\mu_1^2}\cosh\mu_2x\bigg] \nonumber\\
 &&~\nonumber\\
&& + \tilde{P}_S(0,s) \bigg[\frac{v^2}{D^2 \mu_1^2}\cosh\mu_1x + \frac{2w}{D\mu_1^2}\cosh\mu_2 x \bigg]\nonumber\\
&&~\nonumber\\
&&+ \tilde{P}_A(0,s)\bigg[-\frac{vD}{v^2+wD}\big(\mu_0\sinh\mu_0x + \frac{v^2}{D^2\mu_1}\sinh\mu_1x \big) + \frac{2 v\mu_2}{D\mu_1^2}\sinh\mu_2x\bigg]\nonumber\\
&&~\nonumber\\
&& +\theta(x-x_0) \bigg[\frac{1}{2(v^2+wD)}\bigg(\frac{v^2+2wD}{D\mu_0}\sinh\mu_0(x-x_0)+\frac{wv^2 \sinh\mu_1(x-x_0)}{\mu_1(v^2+2wD)}\bigg)\nonumber\\
&&~~~~~~~~~~~~~~~~~~~~~~~~~~~ - \frac{w \sinh\mu_2(x-x_0)}{2\mu_2(v^2+2wD)} \bigg].
\end{eqnarray}
We note that, in the last expression for $\tilde{P}(x,s)$ there are exponentially
growing terms, $e^{\mu_ix}$, that must identically vanish to satisfy the boundary condition at infinity, i.e., $\tilde{P}(x\rightarrow\infty,s)=0$. Since $e^{\mu_ix}$ are linearly
independent for different $i$, we consequently have three equations each containing the three yet undermined factors 
$\tilde{P}_{||}(0,s),~\tilde{P}_S(0,s),~\tilde{P}_A(0,s)$. Solving for them we obtain, \textit{in the small $s$ limit},
\begin{eqnarray}
 &&\tilde{P}_{||}(0,s)=\frac{w}{v\mu_0}\frac{v^2+wD}{2D^2\mu_1(v+wD\frac{\mu_1}{v\mu_0})}\frac{e^{-\mu_2x_0}}{\mu_2} \nonumber\\
 &&\tilde{P}_S(0,s)=\frac{v\mu_1+\frac{w^2}{v\mu_1}}{2D\mu_1(v+wD\frac{\mu_1}{v\mu_0})}\frac{e^{-\mu_2x_0}}{\mu_2} \nonumber\\
 &&\tilde{P}_{A}(0,s)=\frac{v^2+wD}{2D^2\mu_1(v+wD\frac{\mu_1}{v\mu_0})}\frac{e^{-\mu_2x_0}}{\mu_2}\nonumber
\end{eqnarray}
~~~~~ and finally,
\begin{equation}
 \tilde{P}(x,s)=\frac{1}{4D_e}\bigg(\frac{e^{-\mu_2|x-x_0|}}{\mu_2} + \frac{e^{-\mu_2(x+x_0)}}{\mu_2}\bigg) +\frac{v^2 \big(w\mu_1 e^{-\mu_0 x}+\frac{v^2\mu_0}{D}e^{-\mu_1x} \big)}{2D^2\mu_1^2(wD\mu_1+v^2\mu_0)}\frac{e^{-\mu_2 x_0}}{\mu_2}.
\end{equation}
Carrying out the inverse Laplace transform in $s$ we arrive at the equation (\ref{inflinesoln}) in the text.

\end{document}